\documentclass[11pt]{article}

\usepackage[latin1,applemac]{inputenc}
\usepackage[cyr]{aeguill}
\usepackage{authblk}
\usepackage[T1]{fontenc}
\usepackage{hyperref}
\usepackage{amsfonts} 
\usepackage{amssymb}
\usepackage{amsmath}
\usepackage{latexsym}
\usepackage{graphicx}
\usepackage{caption}
\usepackage{verbatimbox}
\usepackage{fancyvrb}
\usepackage{placeins}
\usepackage{multirow}
\usepackage[noadjust]{cite}

\renewcommand{\(}{\left(}
\renewcommand{\)}{\right)}
\usepackage{simplewick}

\usepackage{comment}

\makeatletter
\newcommand\thefontsize[1]{{#1 The current font size is: \f@size pt\par}}
\makeatother

\numberwithin{equation}{section}

\title{Decoupled algorithm for the multicomponent potential theory of adsorption of gas mixtures} 
\author{Rapha\"el Gervais Lavoie\footnote{Email: \url{raphael.gervaislavoie@uqtr.ca}}}
\author{Jean Hamelin}
\author{Pierre B\'enard}

\affil{Institut de recherche sur l'hydrog\`ene, Universit\'e du Qu\'ebec \`a Trois-Rivi\`eres, \break
D\'epartement de chimie, biochimie et physique, Universit\'e du Qu\'ebec \`a Trois-Rivi\`eres.}

\begin{document}
\maketitle

\abstract{In this paper, we present a new implementation of the Multicomponent Potential Theory of Adsorption model. The proposed interpretation makes a clear distinction between parameters depending on the adsorbent from those depending on the adsorbate, which leads to a better understanding of the parameters signification. The interdependence between pure isotherms is eliminated, which means that each component can be individually finely adjusted. This new approach was tested against 14 datasets for a total of 510 experimental mixture adsorption data of CH$_4$, CO$_2$, N$_2$, H$_2$, O$_2$, H$_2$S, C$_2$H$_6$, C$_3$H$_6$ and C$_3$H$_8$ on activated carbons, MOF and zeolites. A slight improvement of 4.67\% on excess adsorption predictions was found, leading to an overall average error of 6.97\% for total excess adsorption and 15.30\% for combined mixtures and components excess adsorption predictions.}


\section{Introduction}

In the standard definition of the Multicomponent Potential Theory of Adsorption model (MPTA), some fitting parameters are interdependent, which requires the simultaneous fitting of pure isotherms. This situation results from the choice of minimizing the number of adjustable parameters of the model. The proposed reinterpretation of the model eliminates this interdependence by introducing new adjustable parameters, specific to each gas component, which ultimately, simplify the adjustment and understanding of the model. Both approaches were tested against 14 different experimental datasets from the literature \cite{Sudibandriyo2003, Dreisbach1999, Schell2012, he2004adsorption, Klouste2018, kloutse2018hydrogen, grande2003propane, talu1996, bakhtyari2014, mofarahi2015experimental, hefti2015adsorption, belmabkhout2007, talu1986}. The datasets include 510 individual mixture adsorption measurements, in which 72 are ternary mixtures adsorption. The fluids considered are CH$_4$, CO$_2$, N$_2$, H$_2$, O$_2$, H$_2$S, C$_2$H$_6$, C$_3$H$_6$ and C$_3$H$_8$. The adsorbent materials are activated carbons (Filtrasorb-400, Norit-R1, AP-360, BPL), metal-organic frameworks (MOF-5, CuBTC), and zeolites (4A, 5A, 13X, ZSM-5, Mordenite). The experiments were performed both volumetrically and gravimetrically at temperatures ranging from 297K to 473K. 


\bigskip
\subsection{Pure gas MPTA model}

When talking about adsorption, it is useful to define the \emph{bulk phase} as the region far from the adsorbent where the fluid is unaffected by the adsorbent material. Conversely, the \emph{adsorbed phase} will represent the region near the surface where the fluid is significantly affected by the presence of the adsorbent material.

The potential theory of adsorption (PTA) is a two-parameter thermodynamic model developed by Shapiro, and Stenby \cite{Shapiro1998} based on the pore filling approach of Polanyi's theory of adsorption \cite{Polanyi1963}. The PTA model was generalized to MPTA for gas mixtures adsorption by Shapiro, Stenby and Monsalvo \cite{Shapiro1998, Monsalvo2007}. The MPTA model supposes that the fluid\nobreakdash--surface interaction is entirely described by a local potential field $\varepsilon$, generated by the surface (\cite{Monsalvo2007a, monsalvo2009modeling}). A common choice for this purpose is the Dubinin--Radushkevich--Astakhov (\cite{Dubinin1989, dundar2014a, dundar2014}) potential, given by
\begin{align}
\varepsilon(z)=\left\{
\begin{array}{ll}
\varepsilon_0\(\ln\frac{z_0}{z}\)^{1/\beta} & \text{if } 0\leq z\leq z_0, \\
0 & \text{if }z> z_0,
\end{array}
\right.
\end{align}
where $\varepsilon_0$ and $z_0$ are the \emph{characteristic energy of adsorption} and the \emph{limiting micropore volume}, respectively. $\beta$ is a parameter which is usually interpreted as a quantification of the heterogeneity of the adsorbent \cite{stoeckli1998recent, terzyk2002kind}. Usually, for activated carbon, the parameter $\beta$ is set to $2$, while $\varepsilon_0$ and $z_0$ are determined by fitting the model to experimental data (see \cite{lavoie2018numerical} for details). The ratio $z/z_0$ represents the fraction of the microporous volume associated with an energy $\varepsilon(z)$. 

The MPTA model is defined by \cite{myers2002thermodynamics, Shapiro1998}
\begin{align}
\mu_B\(T,\rho_B\) = \mu_{Ad}\(T,\rho_{Ad}\) - \varepsilon, \label{mu}
\end{align}
where $\mu_{B}$ and $\rho_B$ are respectively the chemical potentials and the fluid density in the bulk phase, while $\mu_{Ad}$ and $\rho_{Ad}$ are the locals chemical potential and fluid density in the adsorbed phase. The bulk phase properties are assumed to be constant while the adsorbed phase properties vary with position \cite{Shapiro1998}. Using Eq. \eqref{mu}, the adsorbed phase's local thermodynamic properties are uniquely determined from properties of the bulk phase and the values of the parameter $z_0$, $\varepsilon_0$ and $\beta$ through the potential $\varepsilon$. Correspondance between gas pressure and density is carried out through an equation of state. The Nist REFPROP is used here for density and chemical potential calculations \cite{Refprop, lavoie2018numerical}. In the following, we will omit the temperature dependence since $T$ is assumed to be constant.


\bigskip
Eq. \eqref{mu} is inverted to obtain $\rho_{Ad}(z)$ from the chemical potentials. The (Gibbs) excess adsorption $N_{ex}$ (which is what is experimentally measured) is then calculated from
\begin{align}
N_{ex}(\rho_B) = \int_0^{z_0} \(\rho_{Ad}(z) - \rho_B\)dz.\label{Nex}
\end{align}

Optimal values for the fittings parameters are obtained by minimizing the difference $N_{ex}(\rho_B)-N_{ex}^{exp}(\rho_B)$ for pure gases isotherms. The fitting is performed by a Python implemented Levenberg-Marquardt algorithm \cite{newville2016lmfit}. 

\bigskip
\subsection{Gas Mixtures}

For gas mixtures with $M$ components, the simplest approach is to consider that each fluid component $i$ is affected by its own surface potential
\begin{align}
\varepsilon^i(z)=\left\{
\begin{array}{ll}
\varepsilon^i_0\(\ln\frac{z_0}{z}\)^{1/\beta} & \text{if } 0\leq z\leq z_0, \\
0 & \text{if }z> z_0,
\end{array}
\right. &&i=1\dots M,
\end{align}
where $\varepsilon^i_0$ refers to a given component. The parameters $z_0$ and $\beta$ are generally assumed to be common to all mixture components \cite{naseri2014modeling}. Eq. \eqref{mu} now becomes a non-linear coupled system of $M$ equations
\begin{align}
\mu_B^i(\rho_B,x_B^i) + \varepsilon^i(z) - \mu_{Ad}^i(\rho_{Ad}(z),x_{Ad}^i(z)) &= 0, && i=1\dots M,\label{chempot}
\end{align}
in which $x^i$ is the molar fraction of a component $i$ of the mixture. Due to the selectivity of the adsorbent material, the local molar fraction $x^i_{Ad}(z)$ will vary in the adsorbed phase, whereas the molar fraction of the bulk phase $x_B^i$ is constant. Here again, the mixture densities are obtained from pressure measurements, mixture molar fraction, and the REFPROP software.

Equations \eqref{chempot} are solved for $\rho_{Ad}(z)$ and $x_{Ad}^i(z)$. The excess (Gibbs) adsorption of each component in the mixture is obtained from
\begin{align}
N_{ex}^i(\rho_B) = \int_0^{z_0} \rho_{Ad}(z)x_{Ad}^i(z)dz - \rho_Bx_B^i z_0, && i=1\dots M.\label{Nex_i}
\end{align}
Finally, the total adsorbed amount is the sum of the contributions of each component
\begin{align}
N_{ex}(\rho_B) = \sum_{i=1}^M N_{ex}^i(\rho_B).
\end{align}

A key feature of the MPTA model is that the fitting parameters $\varepsilon_i$ and $z_0$ (and possibly $\beta$) are solely obtained from \emph{pure gas adsorption isotherms} in order to predict multicomponent adsorption \cite{Shapiro1998, Monsalvo2007}.

\section{Independent $z_0$ and $\beta$ parameters}

Using unique values of $z_0$ and $\beta$ for all fluids components is generally justified by the fact that those parameters are mostly properties of the adsorbent material. Moreover, this allows the reduction of the fitting parameters to $M+1$ (or $M+2$ if $\beta$ is also fitted).\\

However, there are some disadvantages to this approach. Firstly, all the pure gases must be refit each time that a single component is modified. For example, if we consider a binary mixture of gas A and B, the model must be simultaneously fitted on pure isotherms for gas A and B to obtain $\varepsilon_0^A$, $\varepsilon_0^B$, $z_0$ and $\beta$. Now, if a new mixture of gas A and C is considered, parameters $\varepsilon_0^A$, $z_0$ and $\beta$ cannot be reused. The model must be fit anew using the A and C isotherms. Since $\varepsilon_0^i$ and $z_0$ change every time a component of the mixture is changed, the interpretation of those parameters as \emph{characteristic energy of adsorption} of component $i$ and \emph{limiting micropore volume} become less clear. Indeed, it is expected that at least the characteristic energy of adsorption is constant for the pure adsorption of a pair adsorbate\nobreakdash--adsorbent. This is not the case in the conventional MPTA approach.\\

Secondly, physically speaking, any interaction is characterized by it \emph{strength} and it \emph{range}, as so for the fluid\nobreakdash--surface potential $\varepsilon$. For the sake of the discussion, let us consider the simple graphite adsorbent structure where the surface is essentially constituted of isotropic 2D carbon planes. In that case, the microporous volume $z$ is just a specific surface area times a distance to the surface. From the nearly crystalline structure of the graphite, we can infer that this specific surface area is constant, leaving $z$ being essentially a variable of the distance to the surface. This implies that $z_0$ will also be the product of the same characteristic surface times a characteristic distance to the surface. Any characteristic distance to a surface surely represents a range of interaction, and then, functionally speaking, this means that $z_0$ represents the range of the fluid\nobreakdash--surface interaction. This leaves $\varepsilon_0$ representing the strength of the interaction. 

For disorganized adsorbent structures, the situation is more complicated, but $z$ still can be interpreted as a measure of the distance to the surface time a specific surface. However, this time, the specific surface is given by some complicated geometrical average of the porous surface.

The upshot is that $z_0$ is linked to the range of the interaction, and then, it makes much more sense to consider different $z_0$ for different pure gases rather than the same $z_0$ for all gases. \\

Finally, in the perspective of complex mixtures with many components, it will be even more challenging to fit all these pure isotherms simultaneously rather than fitting each component individually.\\

From all those considerations, individual values of $z_0$ and $\beta$ can be introduced from minor modifications of the fluid\nobreakdash--surface potential which now reads
\begin{align}
\varepsilon^i(z)=\left\{
\begin{array}{ll}
\varepsilon^i_0\(\ln\frac{z_0^i}{z}\)^{1/\beta^i} & \text{if } 0\leq z\leq z_0^i, \\
0 & \text{if }z> z_0^i,
\end{array}
\right. &&i=1\dots M.\label{DRA-mod}
\end{align}
Now, $\epsilon_0^i$, $z_0^i$ and $\beta^i$ are parameters specific to pure gas $i$. The modified potential \eqref{DRA-mod} induce no modification to the system of equation \eqref{chempot}. \\

For excess adsorption, the situation is more complicated. It was said earlier that the adsorbed phase is the region where the fluid is affected by the presence of the surface. This definition now needs to be clarified and extended to the indirect effects of other gases components. Indeed, let us consider the region $z_0^i<z\leq z_0^j$. In that region, the surface potential $\varepsilon^i(z)=0$ since $z>z_0^i$, which seems to indicated that the gas $i$ is unaffected by the presence of the adsorbent. However, the component $j$ will be affected by the presence of the adsorbent in that region since $\varepsilon^j(z)\neq0$ as $z\leq z_0^j$. However, the fact that the component $j$ is affected by the adsorbent will modify its local molar fraction $x_{Ad}^j(z)$. Since $\sum_i x_{Ad}^i=1$, local molar fractions are not independent and then, $x_{Ad}^i$ will be affected indirectly by the adsorption of component $j$. \\

\begin{figure}[!htb]
\centering
\includegraphics[width=12cm]{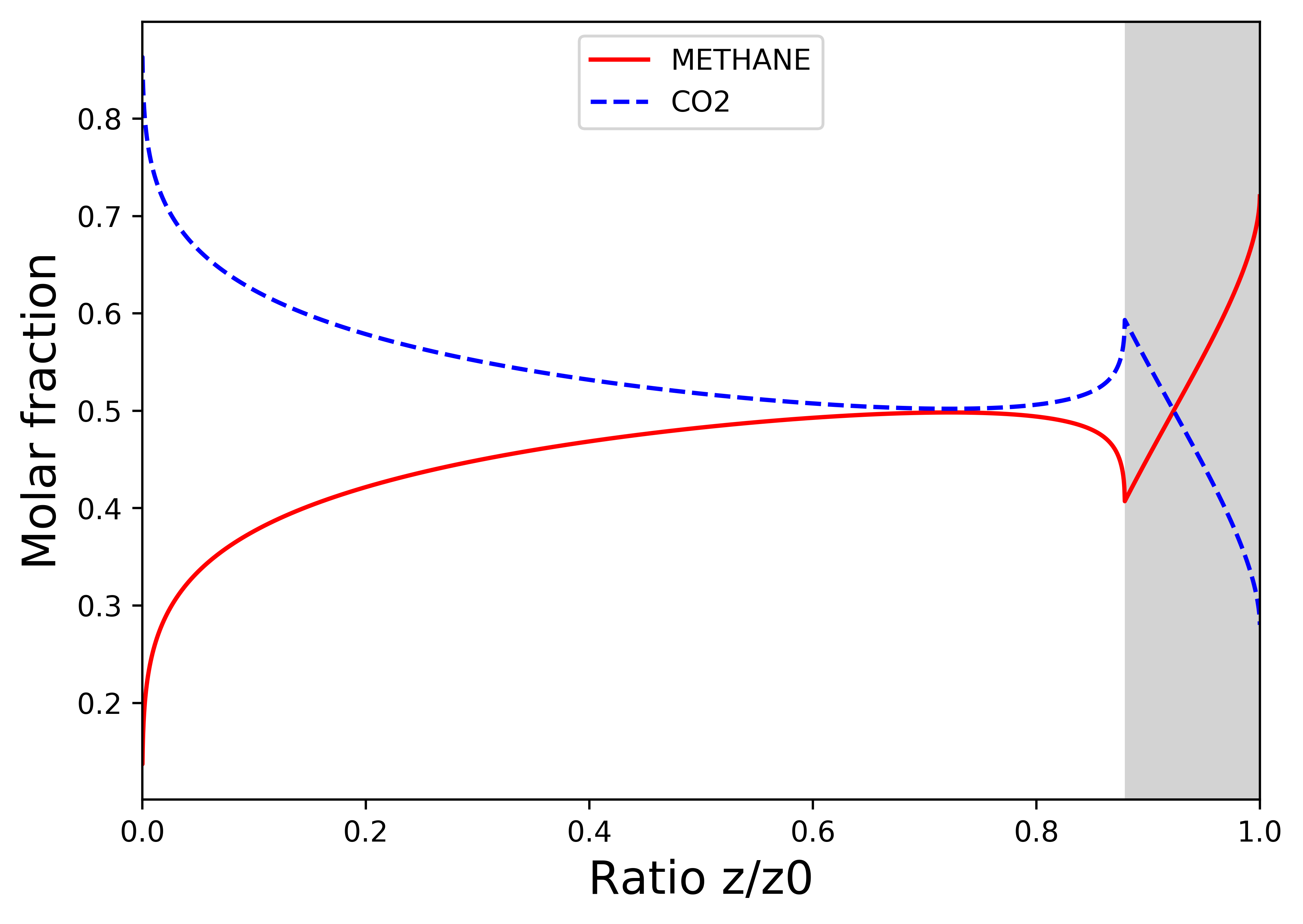}
\caption{\label{MolarFraction1}\textsl{\small Adsorbed phase molar fraction given by the new model for a 72\% CH$_4$ / 28\% CO$_2$ mixture (bulk pressure of 8.3 MPa) on Calgon F-400 activated carbon at 318.2K. The grey area represent the region were CH$_4$ surface potential vanishes, but not the CO$_2$ ones.}}
\end{figure}

The easiest way to see this is by looking at the molar fraction of component $i$ in the range $z_0^i<z\leq z_0^j$, which would have been constant if component $i$ was not affected at all. Fig. \ref{MolarFraction1} show this situation for a mixture of 72\% CH$_4$/28\% CO$_2$ at bulk pressure of 8.3 MPa and temperature of 318.2 K (experimental data were taken in \cite{Sudibandriyo2003}). In the region $z_0^i<z\leq z_0^j$ (the light grey area), we see that the molar fraction of CH$_4$ vary with $z$ even if the surface potential $\varepsilon^{CH_4}(z)$ vanish in that region. At $z=z_0^i$, the CH$_4$ starts to interact with the surface through non-vanishing $\varepsilon^{CH_4}(z)$, and we observe a change in the comportment of the molar fraction. The sharp variation of the molar fraction at $z=z_0^i$ is obviously not physical. It came from the DRA potential which is not smooth at $z=z_0^i$.\\

\begin{figure}[!htb]
\centering
\includegraphics[width=12cm]{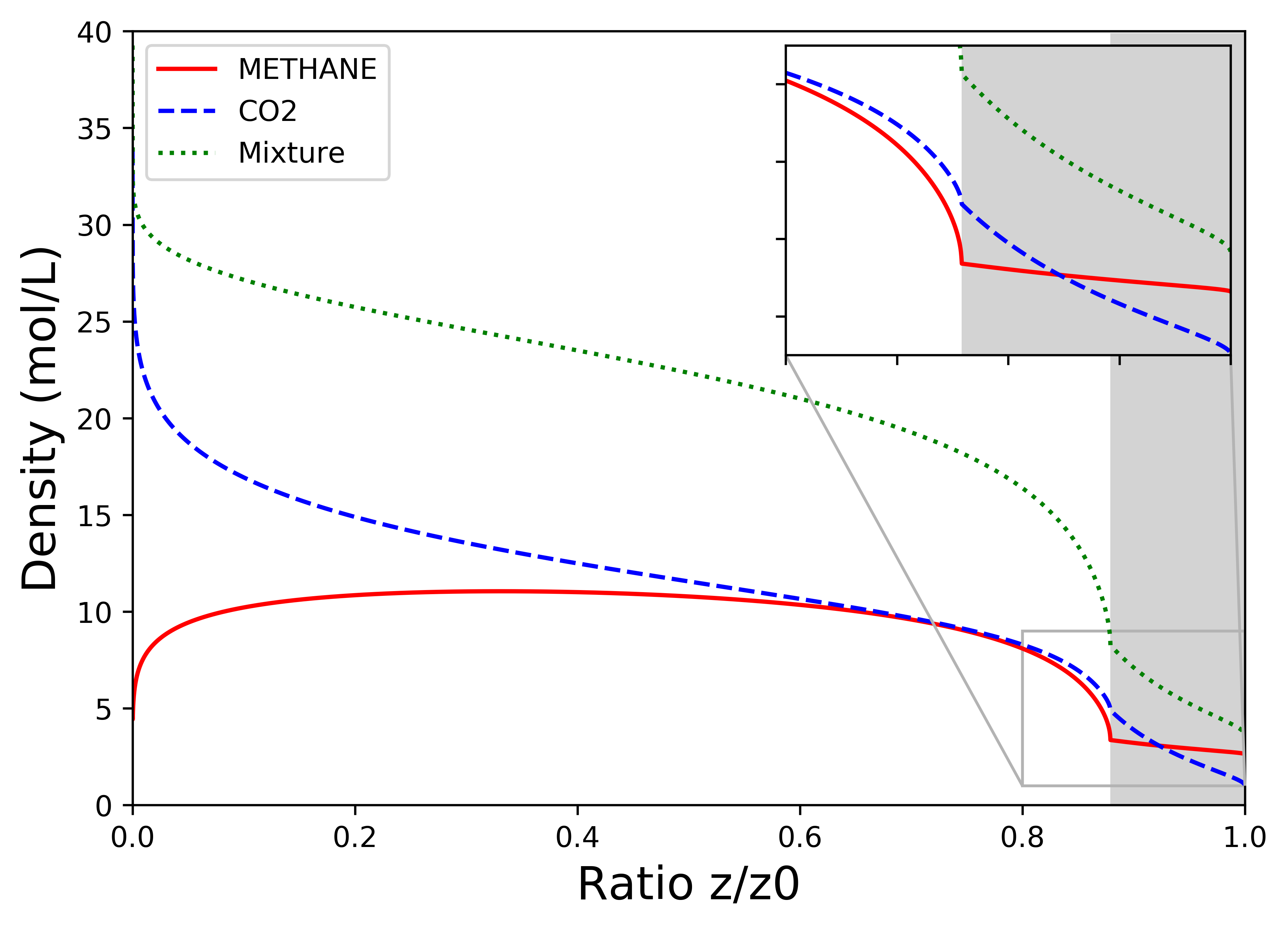}
\caption{\label{DensityProfile1}\textsl{Density profile of the adsorbed phase given by the new model for a 72\% CH$_4$ / 28\% CO$_2$ mixture (bulk\\ pressure of 8.3 MPa) on Calgon F-400 activated carbon at 318.2K. The grey area represent the region were CH$_4$ surface potential vanishes, but not the CO$_2$ ones.}}
\end{figure}

It is also interesting to take a look at the fluid density in that adsorbed phase region. Fig. \ref{DensityProfile1} shows the density profile of the mixture in the same conditions. This figure shows the contribution of each component to the total density, such that the total fluid density is simply the sum of the individual component density. Here again, the sharp variation of fluid density is not physical but is rather an artifact caused by the DRA potential. 

\bigskip
The key point of this discussion is to realize that regardless of the component, the adsorbed phase's fluid properties differ from the ones of the bulk phase for $z<z_0^{max}$. For $z\geq z_0^{max}$, the integral vanishes and then, the excess (Gibbs) adsorption for each component is be given by
\begin{align}
N_{ex}^i(\rho_B) = \int_0^{z_0^{max}} \rho_{Ad}(z)x_{Ad}^i(z)dz - \rho_Bx_B^iz_0^i, && i=1\dots M,
\end{align}
where $z_0^{max}$ is simply the maximum value in $\{z_0^i\}$. The condition that $\varepsilon^i(z)=0$ if $z>z_0^i$ is required in the implementation since the integration now goes from 0 to $z_0^{max}$ for all component. \\ 

It might be slightly confusing that even with individual $z_0^i$, a unique upper limit for the integral emerges. One can then think that this situation argues in favor of a single value of $z_0$ since the integral is ultimately computed over the same range for all the mixture components. However, it is essential to keep in mind that for every component with $z_0^i<z_0^{max}$, the fluid\nobreakdash--surface interaction stop at $z=z_0^i$; The remaining part of the integral (where $z_0^i<z\leq z_0^{max}$) only account for indirect perturbation of the component $i$. In fact, we can see that this part of the integral have a negative contribution to the excess adsorption since the molar fraction of that component $x^i_{Ad}(z)$ is less than in the bulk phase $x_B^i$, and the density of component $i$ is essentially constant.\\

With the proposed reinterpretation, the model now needs to be fitted on $3M$ parameters ($\varepsilon_0^i$, $z_0^i$, and $\beta^i$) instead of $M+2$ parameters ($\varepsilon_0^i$, $z_0$, and $\beta$). However, the new form of the fluid\nobreakdash--surface potential \eqref{DRA-mod} decouples the fitting parameters for each component. In fact, the $3M$ needed parameters split into $M$ individual three parameters fit. Once optimal $\epsilon_0$, $z_0$ and $\beta$ values have been found for a pure gas, there will be no need to refit the model on this gas. Those individual parameters encapsulate all the required information of a pure gas about the fluid\nobreakdash--surface interaction, whatever the mixture considered. In other words, the proposed interpretation make a clear cut between the fluid\nobreakdash--surface interactions, which are governed by the fitting parameters ($\varepsilon_0$, $z_0$ and $\beta$), and the fluid\nobreakdash--fluid interactions, which are entirely governed by the EOS (the REFPROP in our case) as it should be. Moreover, it is quite easier to do $M$ individual three parameters fit than a single $M+2$ parameters fit.\\

Finally, with gas mixtures adsorption, it is quite useful to compare the affinity of the adsorbent's components. This will be done by the use of the \emph{selectivity $S$} of a component over another one. The selectivity of component $i$ over component $j$ is defined as (\cite{kloutse2018hydrogen})
\begin{align}
S_{i/j}=\frac{N_{ex}^ix_B^j}{N_{ex}^jx_B^i}.
\end{align}


\section{Results}

%

To understand the limitation of the model, it is crucial to use accurate experimental data. Whether a volumetric or gravimetric method is used, the variables that are experimentally measured are the total excess adsorption $N_{ex}^{Tot}$ (considering the pressure drop or increase of mass) and the bulk phase molar fraction $x_B$ (generally using gas chromatography). The adsorbed phase molar fraction $x_{ad}$ is then calculated from the initial and equilibrium states, and the components adsorption are calculated from
\begin{align}
N_{ex}^i = N_{ex}^{Tot}x_{ad}^i.
\end{align}
The point here is that both $N_{ex}^{Tot}$ and $x_{ad}^i$ are tainted by experimental uncertainties such that
\begin{align}
\(\delta N_{ex}^i\)^2 =\(\delta N_{ex}^{Tot}x_{ad}^i\)^2 + \(N_{ex}^{Tot}\delta x_{ad}^i\)^2.
\end{align}
Dividing both side by $\(N_{ex}^i\)^2$, we obtain the relative error propagation equation
\begin{align}
\(\frac{\delta N_{ex}^i}{N_{ex}^i}\)^2 &= \(\frac{\delta N_{ex}^{Tot}x_{ad}^i}{N_{ex}^{i}}\)^2 + \(\frac{N_{ex}^{Tot}\delta x_{ad}^i}{N_{ex}^{i}}\)^2.\label{3.3}
\end{align}
Let us focus on the second term. When considering a mixture of different component behavior, it is not uncommon to come across experimental conditions where $N_{ex}^i$ is very small compared to $N_{ex}^{Tot}$. Since $\delta x_{ad}^i$ is not necessarily small enough to compensate for this difference, it is possible to end up with unacceptably large relative uncertainty. To illustrate this, let us consider a case encounter in the dataset where $N_{ex}^{Tot}\sim 6.2$ mmol/g and $x_{ad}^i\sim0.002$. In that particular case, $N_{ex}^i\sim 0.02$ mmol/g, and then, the last term of \eqref{3.3} gives an unacceptable relative uncertainty of $\sim60\%$ on $N_{ex}^i$. From now on, the experimental data with relative uncertainty greater than 25\% will be omitted from the fits. For the experimental dataset with unknown experimental uncertainties, an experimental error of 1\% on $N_{ex}^{Tot}$ and 1\% on the smallest $x_B^i$ will be assumed to evaluate relative uncertainties. Those values are representative of the usual experimental uncertainties and were established from the experimental dataset with given experimental errors.

Table \ref{Table1} shows the considered datasets and gives the mean pure fit error of both standard MPTA and the new interpretation of the model, which will be labeled ``new MPTA" even if this is more of a reinterpretation of the MPTA rather than a new model.

\begin{table}[!htb]
\centering
\footnotesize
\begin{tabular}{lllcc}
						&							&						&\multicolumn{2}{c}{Mean pure fit error}\\\cline{4-5}
Adsorbent									& Adsorbate						& T(K)		& Std MPTA 	& New MPTA \\\hline
AC Calgon F-400 \cite{Sudibandriyo2003} 		& CH$_4$/N$_2$/CO$_2$ 			& 318.2		& 2.82\%		& 1.91\% \\
AC Norit R1 \cite{Dreisbach1999}				& CH$_4$/N$_2$/CO$_2$			& 298		& 3.17\%		& 1.22\% \\
AC AP3-60 \cite{Schell2012} 					& N$_2$/CO$_2$/H$_2$				& 298		& 2.63\%		& 2.83\%  \\
AC BPL \cite{he2004adsorption}				& CH$_4$/C$_2$H$_6$				&297, 301.4	& 5.03\%		& 1.46\% \\
MOF-5 \cite{Klouste2018, kloutse2018hydrogen}	& CH$_4$/N$_2$/H$_2$/CO$_2$		& 297		& 8.91\%		& 3.60\% \\
CuBTC \cite{kloutse2018hydrogen}				& CH$_4$/N$_2$/H$_2$/CO$_2$		& 297		& 2.96\%		& 2.45\% \\
Zeolite-4A \cite{grande2003propane}			& C$_3$H$_8$/C$_3$H$_6$			& 423/473		& 4.94\%		& 2.40\% \\
Zeolite-5A \cite{talu1996}						& O$_2$/N$_2$					& 296		& 2.82\%		& 1.31\% \\
Zeolite-5A \cite{bakhtyari2014}					& CH$_4$/N$_2$					& 303/323		& 6.14\%		& 3.90\% \\
Zeolite-13X \cite{mofarahi2015experimental}		& CH$_4$/N$_2$					& 303/323		& 7.04\%		& 3.36\% \\
Zeolite-13X \cite{hefti2015adsorption}			& CO$_2$/N$_2$					& 298/318		& 3.07\%		& 0.95\% \\
Zeolite-ZSM-5 \cite{hefti2015adsorption}			& CO$_2$/N$_2$					& 298/318		& 3.72\%		& 2.63\% \\
Zeolite-NaX \cite{belmabkhout2007}			& CO$_2$/CO						& 323/373		& 5.46\%		& 2.24\% \\
Zeolite H-Mordenite \cite{talu1986} 				& CO$_2$/H$_2$S/C$_3$H$_8$		& 303		& 7.25\%		& 3.00\% \\
\end{tabular}
\caption{\label{Table1}\textsl{Pure gas mean fit for all the datasets considered.}}
\end{table}


\FloatBarrier
\subsection{Activated Carbon Filtrasorb-400 (Sudibandriyo)}

First, we consider the adsorption of CH$_4$, CO$_2$, N$_2$ and their binary mixtures on activated carbon Filtrasorb-400 (Calgon Carbon Co.) which have a microporous volume of 0.4950 cm$^3$/g, and a BET surface of 850 m$^2$/g (\cite{Sudibandriyo2003}). The measurements were performed volumetrically at 318.2K with pressure up to 13.8 MPa. 
Adsorption of pure gases was carried out twice to guarantee reproducibility. Both runs were used to fit the MPTA model. Overall, the new model underestimates the mixture adsorption by 3.32\%, while the pure isotherms are overestimated by 2\%.

\begin{figure}[!htb]
\centering
\includegraphics[width=.49\textwidth]{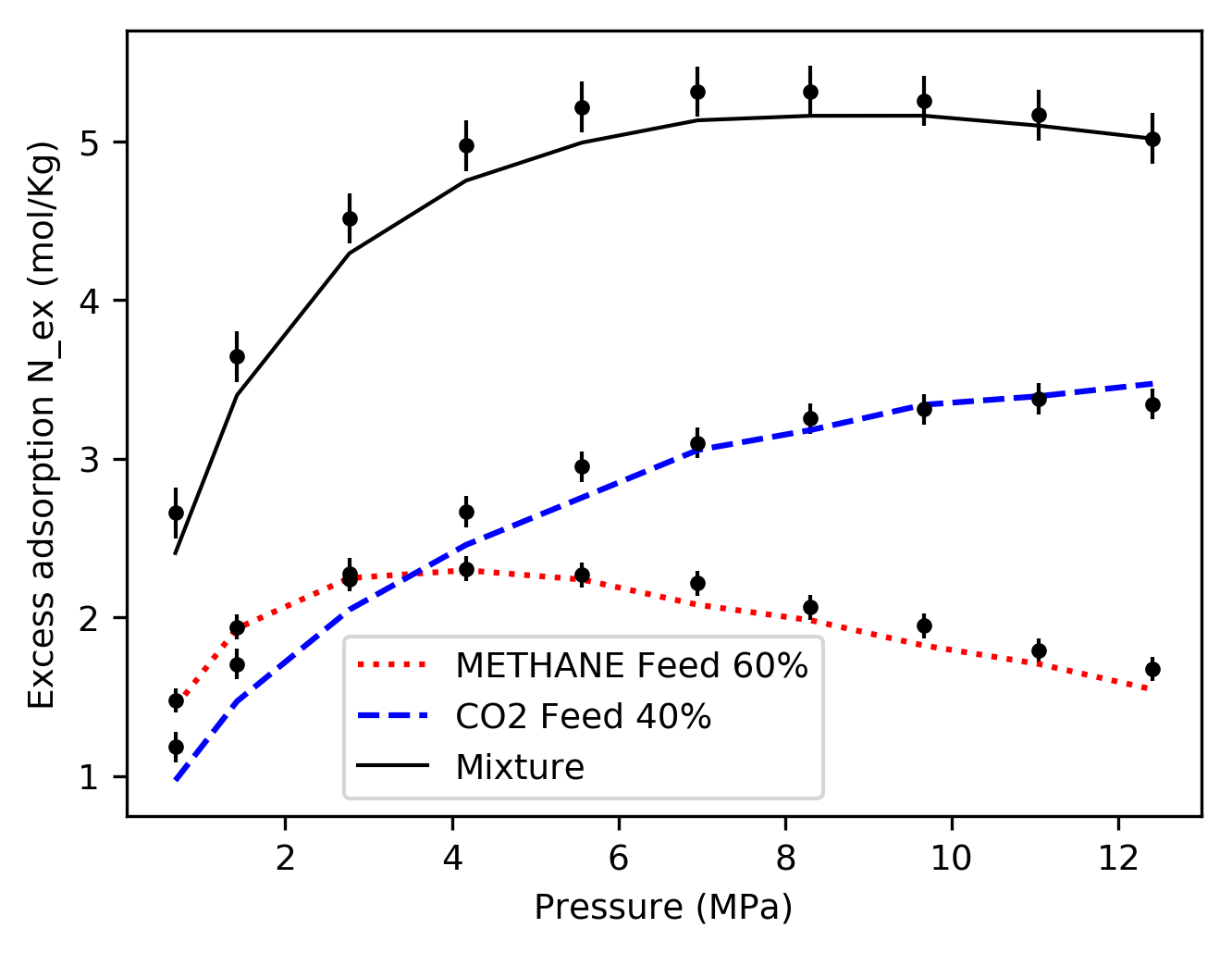}
\includegraphics[width=.5\textwidth]{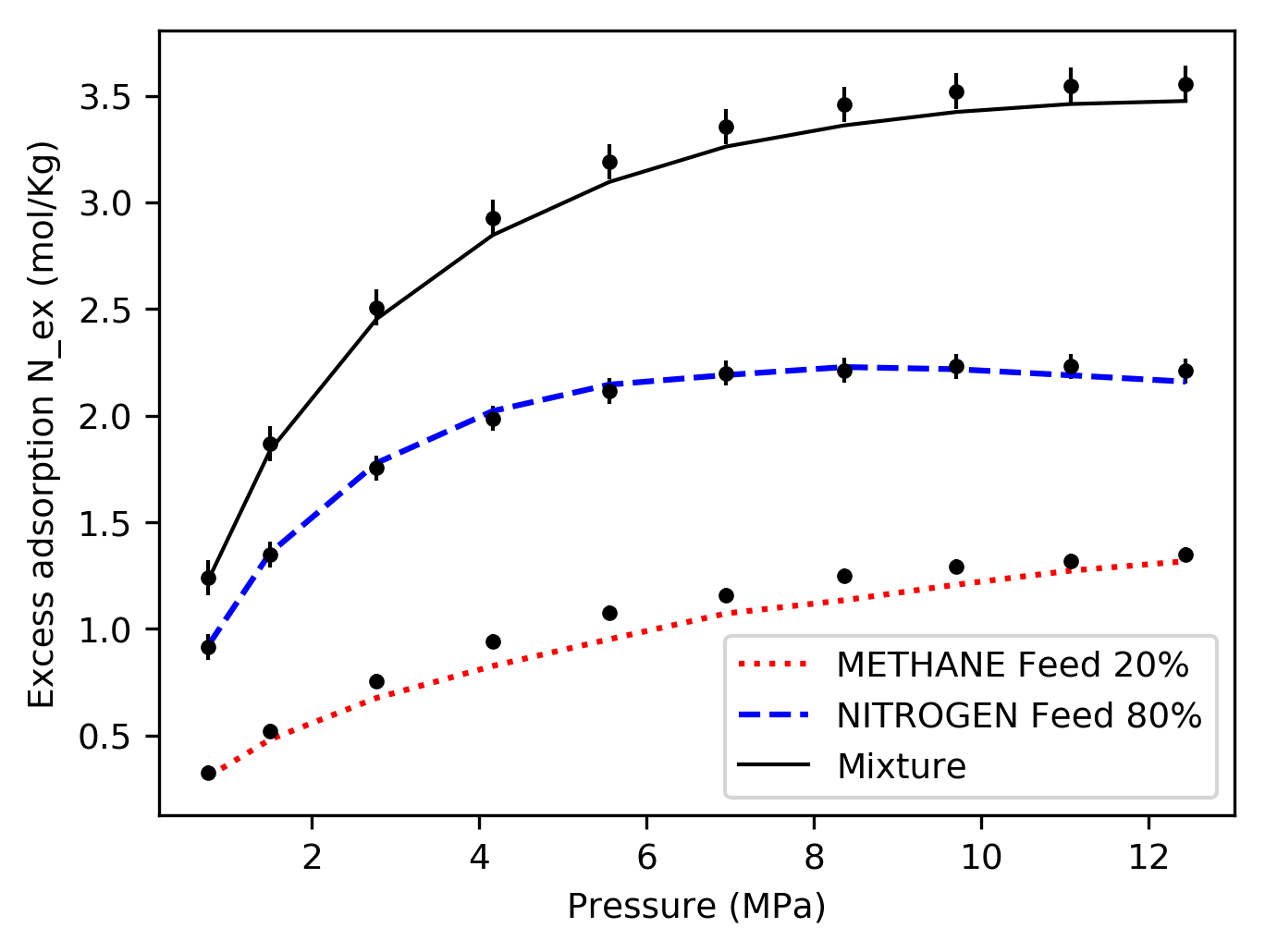}
\includegraphics[width=.5\textwidth]{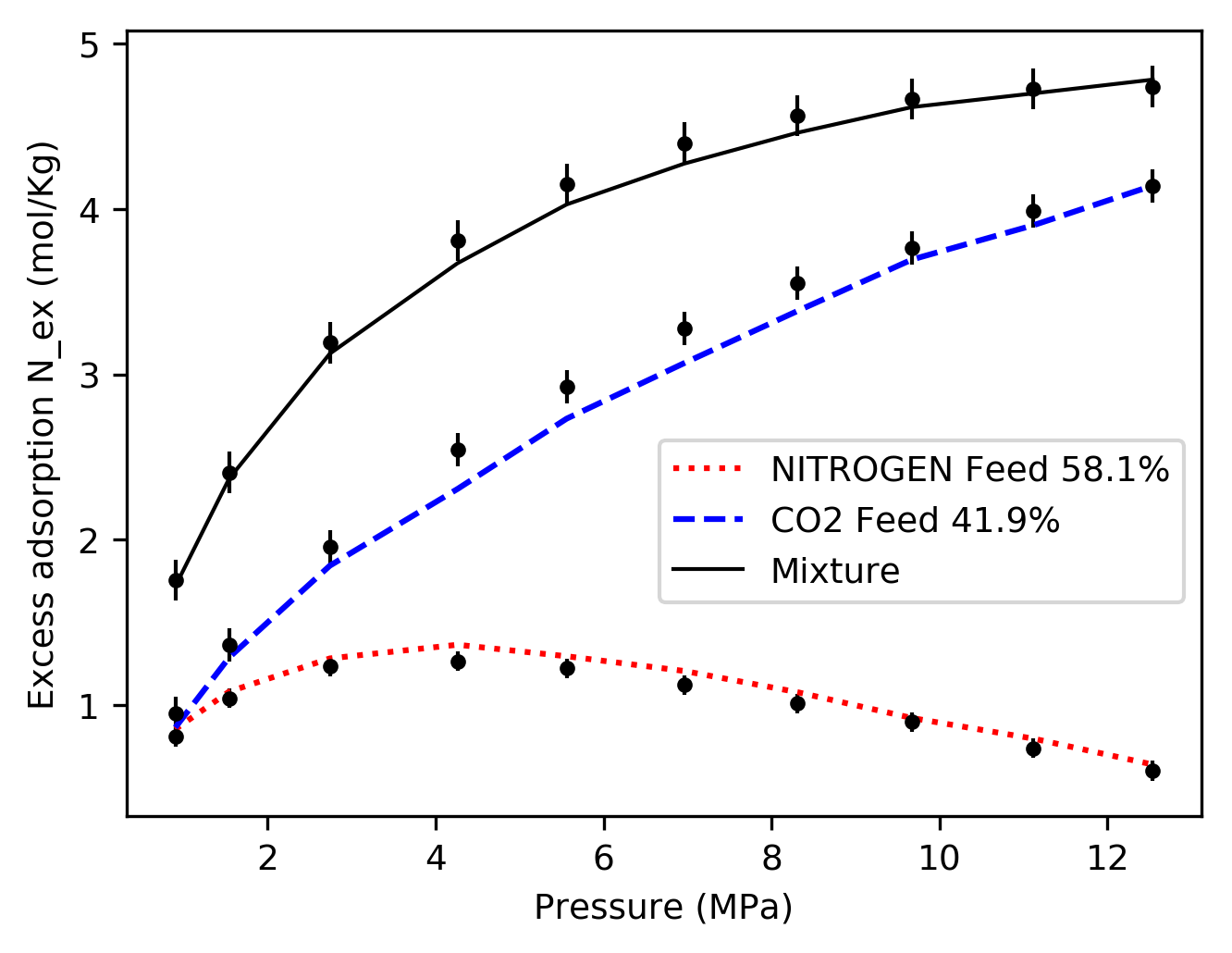}
\caption{\label{Figure3}\textsl{New MPTA model selected examples of binary mixtures on Filtrasorb-400 at 318.2K and pressure up to 13.8 MPa.}}
\end{figure}

Table \ref{Table2} gives the mean error between the prediction of both approaches, while Figure \ref{Figure3} shows some of the new model results.

\begin{table}[!htb]
\centering
\small
\begin{tabular}{llccccc}
							&	& \multicolumn{5}{c}{Mean error (\%)} \\\cline{3-7}
							&	& \multicolumn{2}{c}{Std MPTA} && \multicolumn{2}{c}{New MPTA} \\\cline{3-4}\cline{6-7}
System							&			& N$_{ex}^i$	& Select			&& N$_{ex}^i$	& Select	 \\\hline
CH$_4$/CO$_2$	& CH$_4$ component			& 25.37 	& ---		&& 14.03	& ---	  \\
				& CO$_2$ component			& 6.50	& 52.59	&& 7.88 	& 27.47 \\
				& Mixture						& 2.21 	& ---		&& 3.16 	& ---  \\\hline
CH$_4$/N$_2$	& CH$_4$ component			& 3.30 	& 6.94	&& 7.83	& 11.82 \\
				& N$_2$ component			& 6.39 	& ---		&& 5.40 	& ---	\\
				& Mixture						& 3.36 	& ---		&& 4.13 	& --- 	\\\hline
N$_2$/CO$_2$	& N$_2$ component			& 12.95	& ---		&& 9.98	& --- 	 \\
				& CO$_2$ component			& 6.05 	& 23.70	&& 8.63 	& 21.29 \\
				& Mixture						& 1.49 	& ---		&& 3.30	& --- \\\hline
\multicolumn{2}{l}{Overall error:}					& 7.48	& 27.48	&& 7.12	& 20.02 \\
\multicolumn{2}{l}{Overall increased performance:} 	& --- & ---			&& 4.8	& 27.1 \\\hline
\multicolumn{7}{l}{113 experimental data points.}
\end{tabular}
\caption{\label{Table2}\textsl{Comparison of standard and new MPTA models  on Filtrasorb-400 at 318.2K  and pressure up to 13.8 MPa.}}
\end{table}

\FloatBarrier
\subsection{Activated Carbon Norit R1 (Dreisbach)}

Binary and ternary mixtures of CH$_4$, N$_2$, and CO$_2$ are considered on activated carbon Norit R1 Extra which has a microporous volume of 0.3511 cm$^3$/g, and a BET surface of 1407.3 m$^2$/g (\cite{Dreisbach1999}). The measurements were performed gravimetrically at 298K over a pressure ranging from 93 KPa to 6.077 MPa. Overall, the new model underestimates the mixture's adsorption by 7.68\%, while the pure isotherms are underestimated by 0.82\%.

\begin{figure}[!htb]
\centering
\includegraphics[width=.5\textwidth]{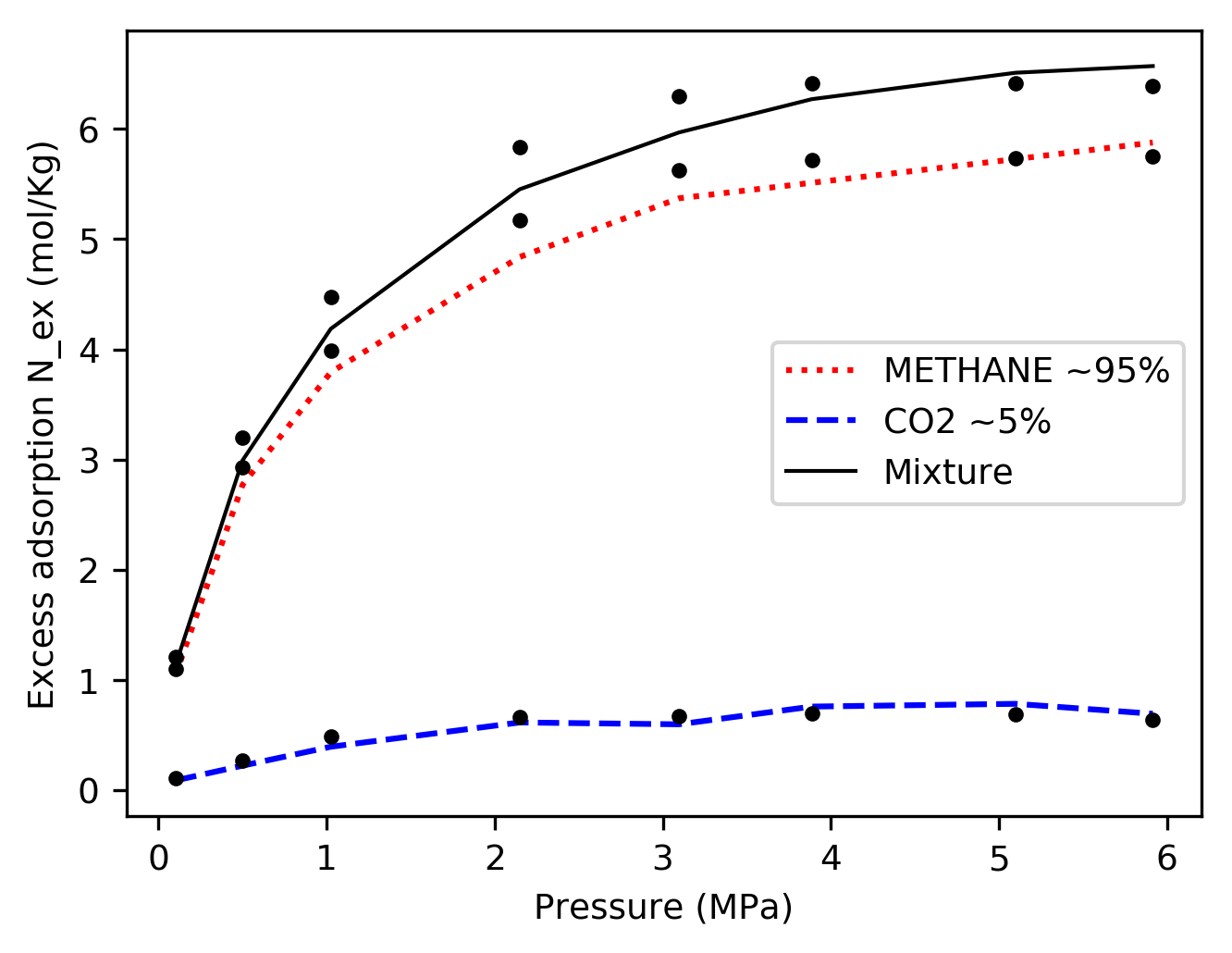}\hfill
\includegraphics[width=.5\textwidth]{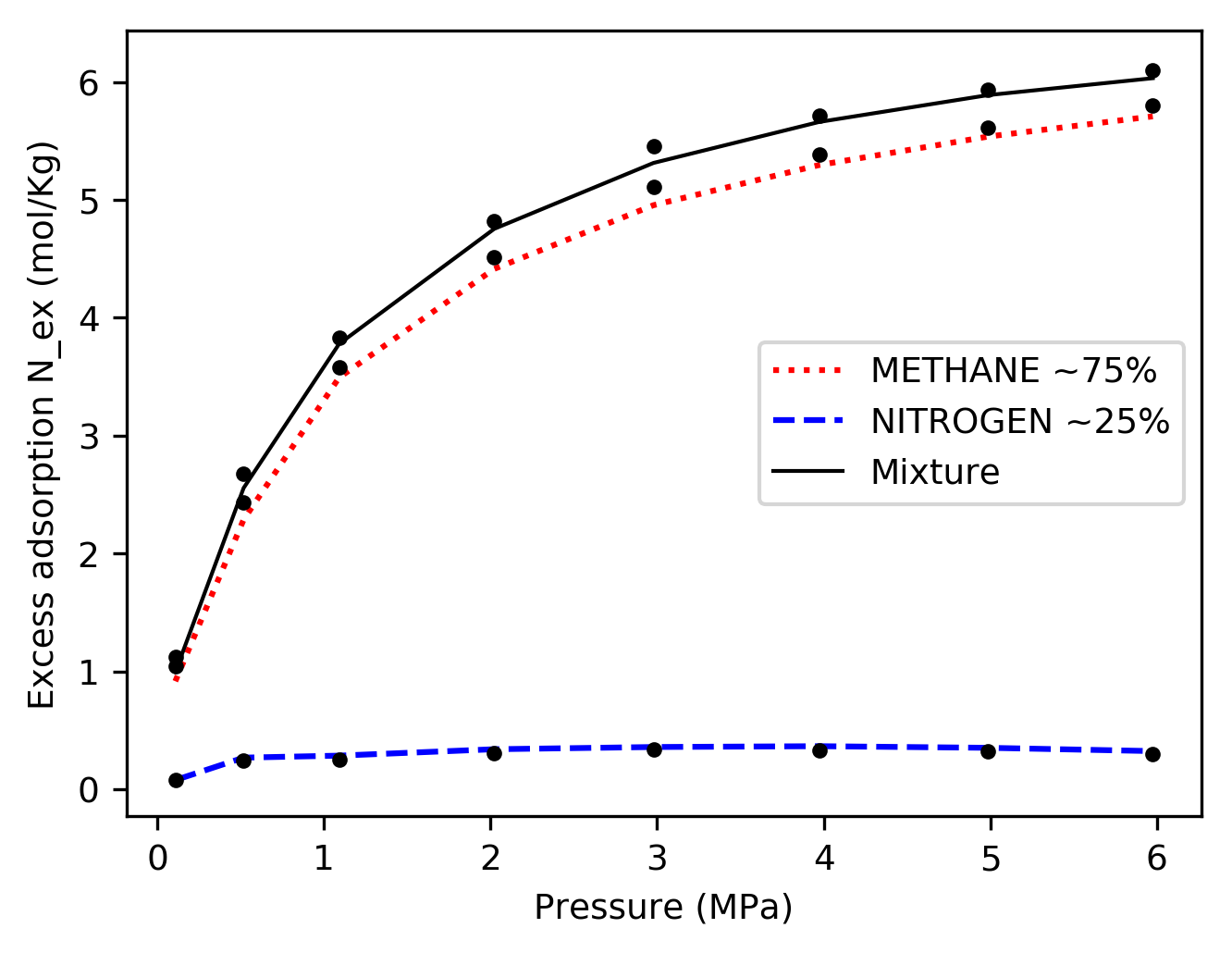}\hfill
\includegraphics[width=.5\textwidth]{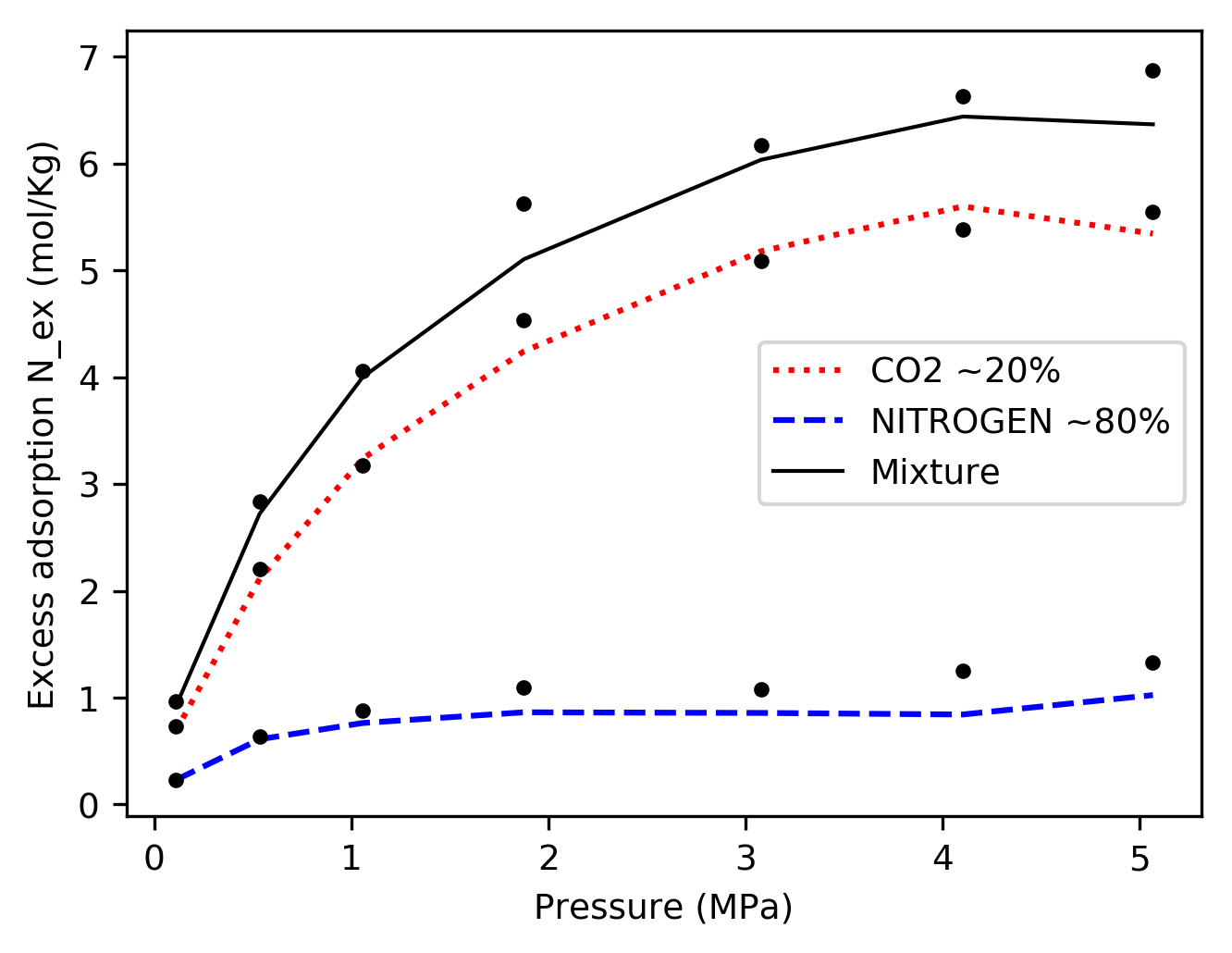}\hfill
\includegraphics[width=.5\textwidth]{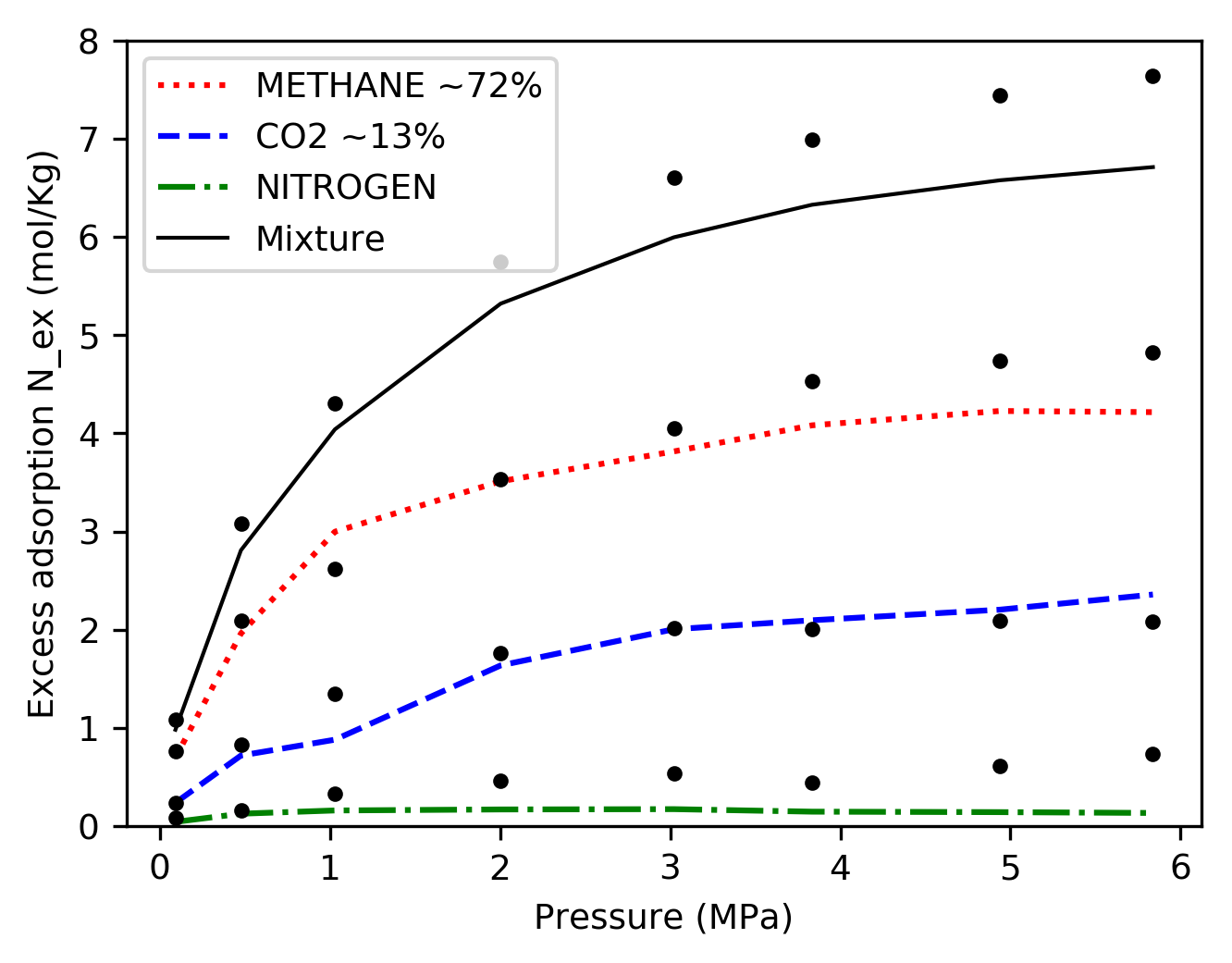}
\caption{\label{Figure4}\textsl{New MPTA model selected examples of binary and ternary mixtures on Norit-R1 at 298 K and pressure up to 6 MPa.}}
\end{figure}

Table \ref{Table3} gives the mean error between the prediction of both approaches, while Figure \ref{Figure4} shows some selected results of the new model.

\begin{table}[!htb]
\centering
\small
\begin{tabular}{llccccc}
							&	& \multicolumn{5}{c}{Mean error (\%)} \\\cline{3-7}
							&	& \multicolumn{2}{c}{Standard MPTA} && \multicolumn{2}{c}{New MPTA} \\\cline{3-4}\cline{6-7}
System				&			& N$_{ex}^i$	& Select			&& N$_{ex}^i$	& Select	 \\\hline
CH$_4$/CO$_2$	& CH$_4$ component		& 36.88 	& ---		&& 36.32		& ---	  \\
				& CO$_2$ component		& 9.07	& 40.97 	&& 7.71 		& 39.83 \\
				& Mixture					& 5.79 	& ---		&& 5.66 		& ---  \\\hline
CH$_4$/N$_2$	& CH$_4$ component		& 7.86 	& 8.93	&& 12.02		& 16.89	 \\
				& N$_2$ component			& 7.00 	& ---		&& 6.13 		& ---	\\
				& Mixture					& 4.93 	& ---		&& 5.36 		& --- 	\\\hline
CO$_2$/N$_2$	& CO$_2$ component		& 4.26	& 27.76	&& 4.40		& 24.72 	 \\
				& N$_2$ component			& 20.45 	& ---		&& 18.30 	& --- \\
				& Mixture					& 3.50 	& ---		&& 5.27		& --- \\\hline
CH$_4$/CO$_2$/N$_2$	
				& CH$_4$ component		& 26.88 	& ---$^\dag$	&& 26.42		& ---$^\dag$	  \\
				& CO$_2$ component		& 16.74 	& ---$^\dag$	&& 14.50 		& ---$^\dag$ \\
				& N$_2$ component			& 57.04 	& ---	 	&& 58.87		& --- 	 \\
				& Mixture					& 10.67 	& ---		&& 11.31		& --- \\\hline
\multicolumn{2}{l}{Overall error:}				& 19.09	& 25.39	&& 19.18		& 27.79 \\
\multicolumn{2}{l}{Overall increased performance:} 	& --- & ---		&& -0.47 		& -9.45  \\\hline
\multicolumn{7}{l}{94 experimental data points.} \\
\multicolumn{7}{l}{\footnotesize $^\dag$ Error on selectivity over 100\% due to large error on the least adsorbed component.}
\end{tabular}
\caption{\label{Table3}\textsl{Comparison of standard and new MPTA models  on Norit-R1 at 298K and pressure up to 6 MPa.}}
\end{table}

\FloatBarrier
\subsection{Activated Carbon AP3-60 (Schell)}

Binary mixtures of CO$_2$, N$_2$ and H$_2$ are considered on activated carbon Envirocarb AP3-60 (Chemviron Carbon) \cite{Schell2012}, which have a BET surface of 1000 m$^2$/g (Chemviron Carbon). The measurements were performed gravimetrically at 298K over a pressure ranging from 400 KPa to 11.86 MPa. Overall, the new model underestimates the mixture adsorption by 9.19\%, while the pure isotherms are underestimated by 1.22\%.

Table \ref{Table4} gives the mean error between the prediction of both approaches.

\begin{table}[!htb]
\centering
\small
\begin{tabular}{llccccc}
							&	& \multicolumn{5}{c}{Mean error (\%)} \\\cline{3-7}
							&	& \multicolumn{2}{c}{Std MPTA} && \multicolumn{2}{c}{New MPTA} \\\cline{3-4}\cline{6-7}
System							&			& N$_{ex}^i$	& Select			&& N$_{ex}^i$	& Select	 \\\hline
CO$_2$/N$_2$	& CO$_2$ component			& 4.47 	& ---			&& 2.40	& ---	  \\
				& N$_2$ component				& 33.45	& 62.57		&& 27.91 	& 52.62 \\
				& Mixture						& 2.60 	& ---			&& 2.64 	& ---  \\\hline
CO$_2$/H$_2$	& CO$_2$ component			& 3.38 	& ---			&& 4.00	& ---	  \\
				& H$_2$ component				& 115.36	& ---$^\dag$	&& 108.44 	& ---$^\dag$ \\
				& Mixture						& 13.30 	& ---			&& 12.47 	& ---  \\\hline				
\multicolumn{2}{l}{Overall error:}					& 34.10	& 62.57		&& 31.68	& 52.62 \\
\multicolumn{2}{l}{Overall increased performance:} 		& --- & ---				&& 7.10	& 15.90 \\\hline
\multicolumn{7}{l}{40 experimental data points.}\\
\multicolumn{7}{l}{\footnotesize $^\dag$ Error on selectivity over 100\% due to large error on the least adsorbed component.}
\end{tabular}
\caption{\label{Table4}\textsl{Comparison of standard and new MPTA models on activated carbon AP3-60 at 298K  and pressure up to 10.8 MPa.}}
\end{table}

\FloatBarrier
\subsection{Activated Carbon BPL (He)}

Binary mixtures of CO$_2$ and C$_2$H$_6$ are considered on activated carbon BPL (Calgon Carbon Co.) \cite{he2004adsorption}, which have a microporous volume of 0,630 cm$^3$/g and a BET surface of 1200 m$^2$/g (\cite{russell1994pore}). The measurements were performed volumetrically at 297K and 301.4K with pressure up to 2.5 MPa. Overall, the new model underestimates the mixture adsorption by 9.76\%, while the pure isotherms are underestimated by 0.06\%.

Table \ref{Table5} gives the mean error between the prediction of both approaches.

\begin{table}[!htb]
\centering
\small
\begin{tabular}{llccccc}
							&	& \multicolumn{5}{c}{Mean error (\%)} \\\cline{3-7}
							&	& \multicolumn{2}{c}{Std MPTA} && \multicolumn{2}{c}{New MPTA} \\\cline{3-4}\cline{6-7}
System							&			& N$_{ex}^i$	& Select			&& N$_{ex}^i$	& Select	 \\\hline
CH$_4$/C$_2$H$_6$ 297K	& CH$_4$ component			& 13.57 	& ---			&& 12.88	& ---	  \\
						& C$_2$H$_6$ component		& 12.98	& 20.92		&& 15.16 	& 20.16 \\
						& Mixture						& 9.04 	& ---			&& 9.64 	& ---  \\\hline
CH$_4$/C$_2$H$_6$ 301.4K& CH$_4$  component			& 24.62 	& ---			&& 23.72	& ---	  \\
						& C$_2$H$_6$ component		& 12.96	& 45.01		&& 13.14 	& 49.26 \\
						& Mixture						& 6.50 	& ---			&& 6.56 	& ---  \\\hline				
\multicolumn{2}{l}{Overall error:}							& 13.43	& 34.30		&& 13.62	& 36.33 \\
\multicolumn{2}{l}{Overall increased performance:} 			& --- & ---				&& -1.41	& -5.92  \\\hline
\multicolumn{7}{l}{54 experimental data points.}
\end{tabular}
\caption{\label{Table5}\textsl{Comparison of standard and new MPTA models  on activated carbon BPL at 297K and 301.4K  and pressure up to 2.5 MPa.}}
\end{table}

\FloatBarrier
\subsection{MOF-5 (Klouste)}

Binary mixtures of CH$_4$, N$_2$ and CO$_2$ and ternary mixture of H$_2$, CH$_4$, N$_2$ and CO$_2$ are considered on metal-organic framework MOF-5 (Basolite C300), which have a microporous volume of 1.31 cm$^3$/g and a BET surface of 3054 m$^2$/g (\cite{Klouste2018, kloutse2018hydrogen}). The measurements were performed volumetrically at 297K with pressure up to 1.5 MPa. Overall, the new model underestimates the mixture adsorption by 9.86\%, while the pure isotherms are underestimated by 1.60\%.

\begin{figure}[!htb]
\centering
\includegraphics[width=.5\textwidth]{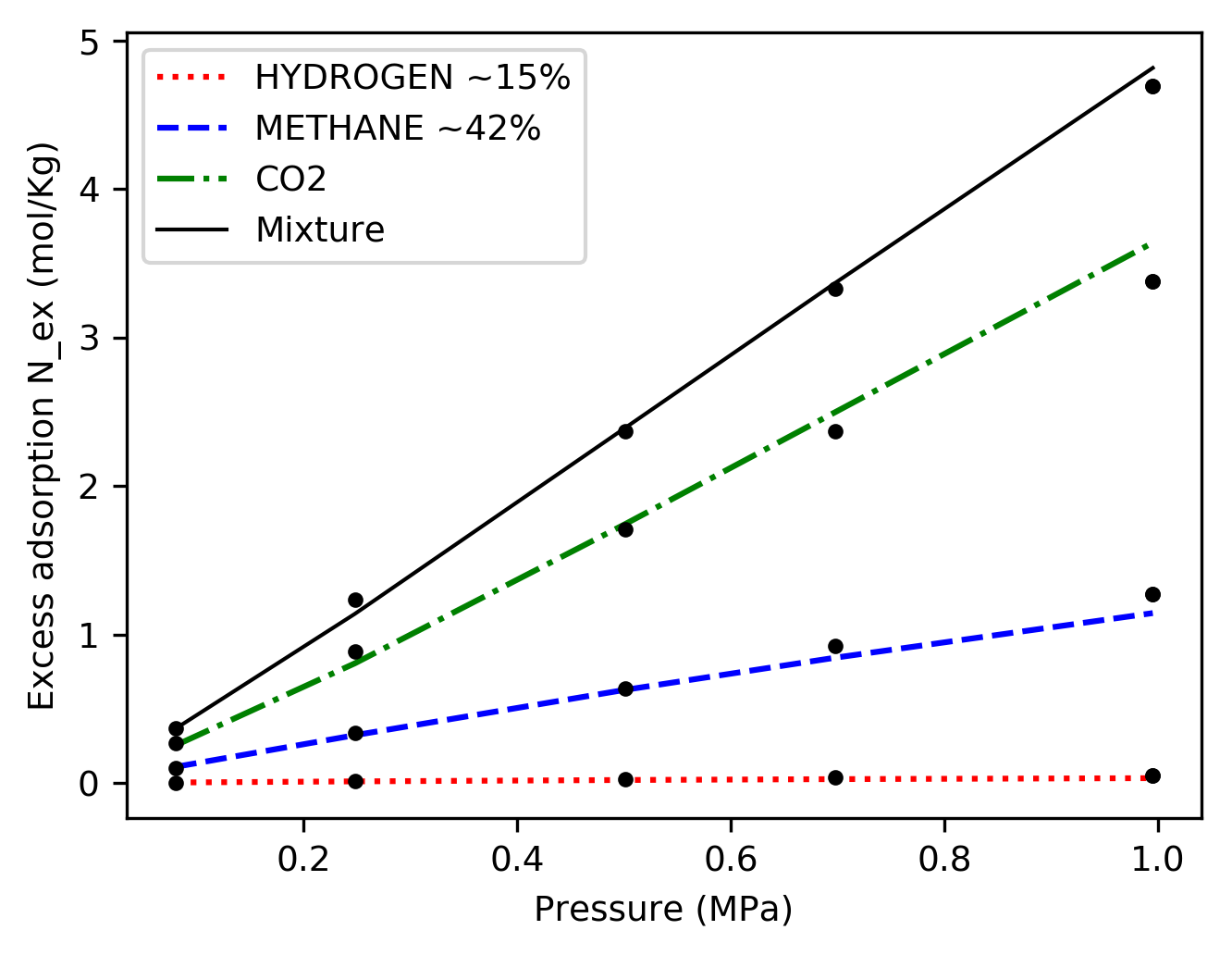}
\caption{\label{Figure5}\textsl{New MPTA model selected examples of ternary mixture on MOF-5 at 297K and pressure up to 1510 KPa.}}
\end{figure}

Table \ref{Table7} gives the mean error between the prediction of both approaches, while Figure \ref{Figure5} shows some selected results of the new model.

\begin{table}[!htb]
\centering
\small
\begin{tabular}{llccccc}
							&	& \multicolumn{5}{c}{Mean error (\%)} \\\cline{3-7}
							&	& \multicolumn{2}{c}{Std MPTA} && \multicolumn{2}{c}{New MPTA} \\\cline{3-4}\cline{6-7}
System				&			& N$_{ex}^i$	& Select		&& N$_{ex}^i$	& Select	 \\\hline
CH$_4$/CO$_2$	& CH$_4$ component			& 23.59	& ---				&& 17.86 	& --- \\
				& CO$_2$ component			& 7.25  	& 39.64			&& 6.67   	& 29.06 \\
				& Mixture						& 5.56 	& ---				&& 4.10 		& --- \\\hline
CH$_4$/N$_2$	& CH$_4$ component			& 3.96 	& 12.76			&& 3.87	 	& 3.96 \\
				& N$_2$ component			& 10.04 	& ---				&& 6.09	  	& --- \\
				& Mixture						& 4.48 	& ---				&& 3.97 	 	& ---	 \\\hline
CO$_2$/N$_2$	& CO$_2$ component			& 5.50 	& 47.77			&& 5.22  		& 33.36 \\
				& N$_2$ component			& 32.83 	& ---				&& 25.90  	& --- \\
				& Mixture						& 7.75 	& ---				&& 7.26   	& --- \\\hline
N$_2$/CH$_4$/CO$_2$	& N$_2$ component		& 28.99 	& ---				&& 20.32  	& --- \\
					& CH$_4$ component		& 22.21	& 10.78$^\dag$	&& 15.79   	& 10.54$^\dag$ \\
					& CO$_2$ component		& 16.36 	& 38.71$^\dag$	&& 15.83   	& 29.81$^\dag$ \\
					& Mixture					& 13.44 	& --- 			&& 10.58   	& --- \\\hline
H$_2$/CH$_4$/CO$_2$	& H$_2$ component		& 7.28 	& ---				&& 17.09	 	& --- \\
					& CH$_4$ component		& 18.77	& 14.16$^\dag$	&& 13.18   	& 7.93$^\dag$ \\
					& CO$_2$ component		& 25.64 	& 33.70$^\dag$	&& 12.12   	& 30.36$^\dag$ \\
					& Mixture					& 21.20 	& --- 			&& 9.07   	& --- \\\hline
H$_2$/N$_2$/CO$_2$	& H$_2$ component		& 19.29 	& ---				&& 26.53  	& --- \\
					& N$_2$ component		& 16.72	& 15.13$^\dag$	&& 5.88 	 	& 29.54$^\dag$ \\
					& CO$_2$ component		& 19.64 	& 37.71$^\dag$	&& 12.05 	& 39.66$^\dag$ \\
					& Mixture					& 13.64 	& --- 			&& 8.52   	& --- \\\hline
\multicolumn{2}{l}{Overall error:}					& 16.43	& 27.64			&& 12.61		& 21.88  \\\hline
\multicolumn{2}{l}{Overall increased performance:} 	& --- 	& ---				&& 23.25		& 20.84   \\\hline
\multicolumn{7}{l}{40 experimental data points.} \\
\multicolumn{7}{l}{\footnotesize $^\dag$ Represent the adsorption selectivity of the component compare to the first component.}
\end{tabular}
\caption{\label{Table7}\textsl{Comparison of standard and new MPTA models on MOF-5 at 297K and pressure from 0 to 1510 KPa.}}
\end{table}

\FloatBarrier
\subsection{CuBTC (Klouste)}

Ternary mixtures of H$_2$, N$_2$ and CO$_2$ are considered on metal-organic framework CuBTC (Basolite Z100H), which have a microporous volume of 0.66 cm$^3$/g and a BET surface of 1556 m$^2$/g (\cite{kloutse2018hydrogen}). The measurements were performed volumetrically at 297K with pressure up to 1 MPa. Overall, the new model underestimates the mixture adsorption by 31.93\%, while the pure isotherms are overestimated by 0.89\%.

Table \ref{Table8} gives the mean error between the prediction of both approaches.

\begin{table}[!htb]
\centering
\small
\begin{tabular}{llccccc}
							&	& \multicolumn{5}{c}{Mean error (\%)} \\\cline{3-7}
							&	& \multicolumn{2}{c}{Std MPTA} && \multicolumn{2}{c}{New MPTA} \\\cline{3-4}\cline{6-7}
System				&			& N$_{ex}^i$	& Select		&& N$_{ex}^i$	& Select	 \\\hline
H$_2$/N$_2$/CO$_2$	& H$_2$ component		& 42.49 	& ---				&& 35.14  	& --- \\
					& N$_2$ component		& 22.08	& 38.92$^\dag$	&& 28.66 	& 11.80$^\dag$ \\
					& CO$_2$ component		& 3.93 	& 83.67$^\dag$	&& 4.18	 	& 63.18$^\dag$ \\
					& Mixture					& 2.90 	& --- 			&& 2.10   	& --- \\\hline
\multicolumn{2}{l}{Overall error:}					& 17.85	& 61.30			&& 17.52		& 37.49  \\\hline
\multicolumn{2}{l}{Overall increased performance:} 	& --- 	& ---				&& 1.85		& 38.84   \\\hline
\multicolumn{7}{l}{3 experimental data points.} \\
\multicolumn{7}{l}{\footnotesize $^\dag$ Represent the adsorption selectivity of the component compare to the first component.}
\end{tabular}
\caption{\label{Table8}\textsl{Comparison of standard and new MPTA models on CuBTC at 297K and pressure from 0 to 1 MPa.}}
\end{table}

\FloatBarrier
\subsection{Zeolite-4A (Grande)}

Binary mixtures of C$_3$H$_8$ and C$_3$H$_6$ are considered on Zeolite-4A \cite{grande2003propane}, which have a microporous volume of 0.2462 cm$^3$/g and a BET surface of 559.13 m$^2$/g (\cite{sowunmi2018dataset}). The measurements were performed volumetrically at 423K and 473K over a pressure ranging from 85 KPa to 145 KPa. Overall, the new model underestimates the mixture adsorption by 1.13\%, while the pure isotherms are overestimated by 0.32\%.

Table \ref{Table9} gives the mean error between the prediction of both approaches.

\begin{table}[!htb]
\centering
\small
\begin{tabular}{llccccc}
							&	& \multicolumn{5}{c}{Mean error (\%)} \\\cline{3-7}
							&	& \multicolumn{2}{c}{Std MPTA} && \multicolumn{2}{c}{New MPTA} \\\cline{3-4}\cline{6-7}
System						&							& N$_{ex}^i$	& S$_{N_2/O_2}$	&& N$_{ex}^i$	& S$_{N_2/O_2}$	 \\\hline
C$_3$H$_8$/C$_3$H$_6$ 423K	& C$_3$H$_8$ component		& 18.65 	& ---		&& 18.46	& --- \\
							& C$_3$H$_6$ component		& 4.03 	& 32.26	&& 3.80	& 30.59 \\
							& Mixture						& 2.89 	& ---		&& 2.94	& --- \\\hline
C$_3$H$_8$/C$_3$H$_6$ 473K	& C$_3$H$_8$ component		& 14.72 	& ---		&& 13.56	& --- \\
							& C$_3$H$_6$ component		& 2.64 	& 15.80	&& 5.86	& 12.15 \\
							& Mixture						& 3.76 	& ---		&& 3.40	& --- \\\hline
\multicolumn{2}{l}{Overall error:}								& 7.84	& 24.66	&& 8.03	& 22.08 \\\hline
\multicolumn{2}{l}{Overall increased performance:} 					& --- 	& ---			&& -2.42	& 10.46  \\\hline
\multicolumn{7}{l}{13 experimental data points.}
\end{tabular}
\caption{\label{Table9}\textsl{Comparison of standard and new MPTA models  on 4A-Zeolite at 423K and 473K and pressure close to 100 KPa}}
\end{table}

\FloatBarrier
\subsection{Zeolite-5A (Talu)}

Binary mixtures of O$_2$ and N$_2$ are considered on a commercial Zeolite-5A (Tosoh Corporation) \cite{talu1996}, which have a microporous volume of 0.198 cm$^3$/g and a BET surface of 561.1 m$^2$/g (\cite{li2006effects}). The measurements were performed volumetrically at 296K over a pressure ranging from 23 KPa to 921 KPa. Overall, the new model underestimates the mixture adsorption by 4.68\%, while the pure isotherms are underestimated by 1.07\%.

\begin{figure}[!htb]
\centering
\includegraphics[width=.49\textwidth]{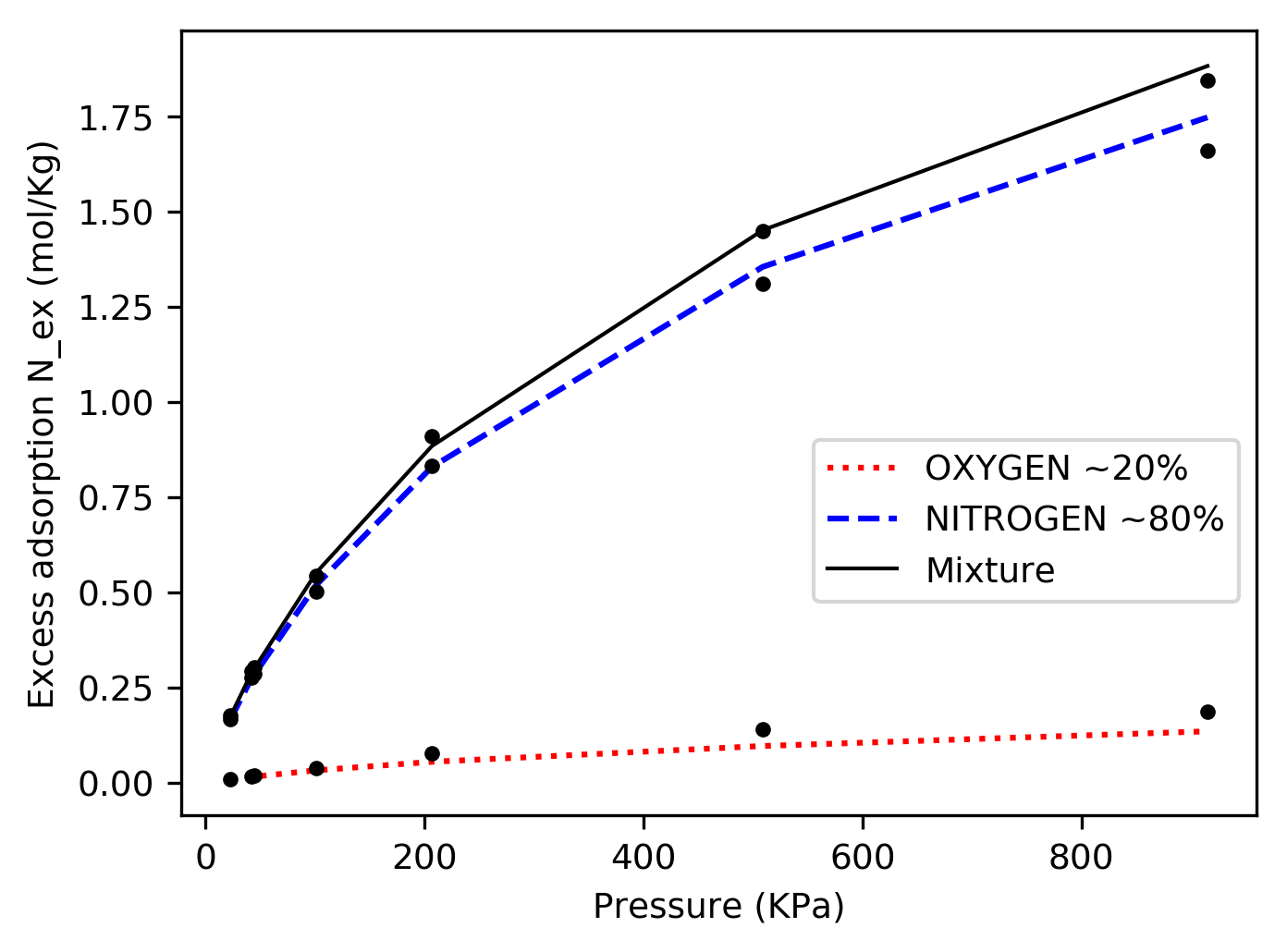}
\includegraphics[width=.49\textwidth]{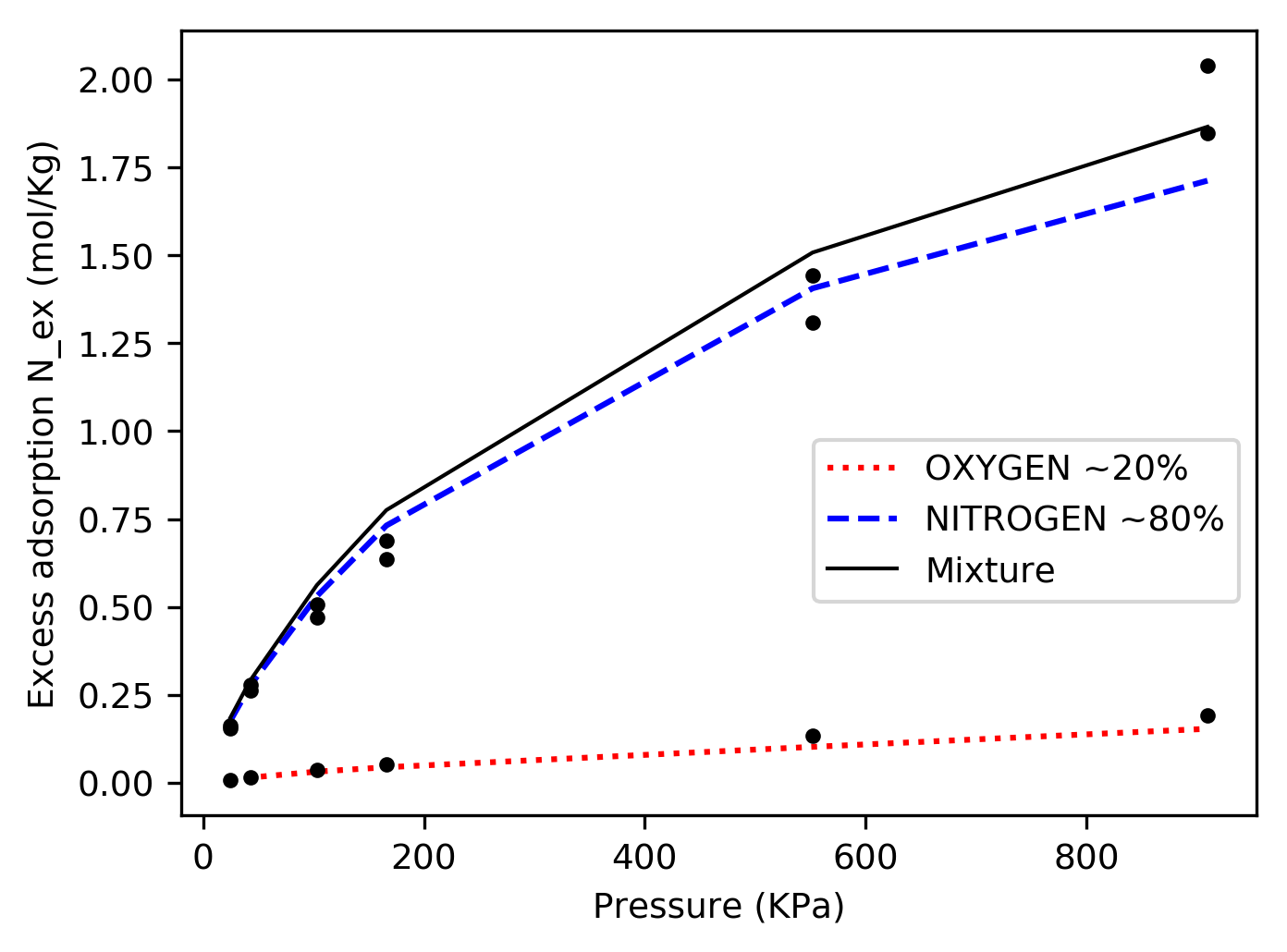}
\caption{\label{Figure6}\textsl{Chosen examples of the new MPTA model for binary mixtures on Zeolite-5A at 296K and pressure up to 921 KPa.}}
\end{figure}

Table \ref{Table10} gives the mean error between the prediction of both approaches, while Figure \ref{Figure6} shows some selected results of the new model.

\begin{table}[!htb]
\centering
\small
\begin{tabular}{llccccc}
							&	& \multicolumn{5}{c}{Mean error (\%)} \\\cline{3-7}
							&	& \multicolumn{2}{c}{Std MPTA} && \multicolumn{2}{c}{New MPTA} \\\cline{3-4}\cline{6-7}
System				&			& N$_{ex}^i$	& S$_{N_2/O_2}$	&& N$_{ex}^i$	& S$_{N_2/O_2}$	 \\\hline
O$_2$/N$_2$ open system	& O$_2$ component					& 24.41 	& ---		&& 21.51	& --- \\
							& N$_2$ component				& 4.48 	& 40.41	&& 2.89	& 33.28 \\
							& Mixture							& 2.39 	& ---		&& 1.62	& --- \\\hline
O$_2$/N$_2$ closed system	& O$_2$ component					& 17.98 	& ---		&& 14.69 	& --- \\
							& N$_2$ component				& 13.70 	& 36.10	&& 10.56	& 26.00 \\
							& Mixture							& 11.98 	& ---		&& 9.27	& ---	 \\\hline
\multicolumn{2}{l}{Overall error:}									& 11.61	& 39.18	&& 9.48	& 31.20 \\\hline
\multicolumn{2}{l}{Overall increased performance:} 					& --- 	& ---		&& 18.3	& 20.3  \\\hline
\multicolumn{7}{l}{21 experimental data points.}
\end{tabular}
\caption{\label{Table10}\textsl{Comparison of standard and new MPTA models  on 5A-Zeolite at 296K and pressure from 23 to 921 KPa}}
\end{table}

\FloatBarrier
\subsection{Zeolite-5A (Bakhtyari)}

Binary mixtures of CH$_4$ and N$_2$ are considered on Zeolite-5A (Zeochem Co.), which have a BET surface of 457--600 m$^2$/g (\cite{bakhtyari2014}). The measurements were performed volumetrically at 303K and 323K over a pressure ranging from 98 KPa to 916 KPa. Overall, the new model overestimates the mixture adsorption by 1.34\%, while the pure isotherms are underestimated by 3.03\%.

Table \ref{Table11} gives the mean error between the prediction of both approaches. 

\begin{table}[!htb]
\centering
\small
\begin{tabular}{llccccc}
							&	& \multicolumn{5}{c}{Mean error (\%)} \\\cline{3-7}
							&	& \multicolumn{2}{c}{Std MPTA} && \multicolumn{2}{c}{New MPTA} \\\cline{3-4}\cline{6-7}
System				&		& N$_{ex}^i$	& S$_{CH_4/N_2}$	&& N$_{ex}^i$	& S$_{CH_4/N_2}$	 \\\hline
CH$_4$/N$_2$ 303K			& CH$_4$ component				& 6.00 	& ---		&& 9.32	& --- \\
							& N$_2$ component				& 6.57 	& 6.22	&& 5.21	& 13.61 \\
							& Mixture							& 5.86 	& ---		&& 5.81	& --- \\\hline
CH$_4$/N$_2$ 323K			& CH$_4$ component				& 4.11 	& ---		&& 3.97 	& --- \\
							& N$_2$ component				& 5.12 	& 4.24	&& 8.37	& 9.58 \\
							& Mixture							& 3.99 	& ---		&& 3.77	& ---	 \\\hline
\multicolumn{2}{l}{Overall error:}									& 5.61	& 5.61	&& 6.35	& 12.37 \\\hline
\multicolumn{2}{l}{Overall increased performance:} 					& --- 	& ---		&& -13.19	& -120.50  \\\hline
\multicolumn{7}{l}{26 experimental data points.}
\end{tabular}
\caption{\label{Table11}\textsl{Comparison of standard and new MPTA models  on Zeolite-5A at 303K and 323K, and pressure from 98 to 916 KPa}}
\end{table}

\FloatBarrier
\subsection{Zeolite-13X (Bakhtyari)}

Binary mixtures of CH$_4$ and N$_2$ are considered on Zeolite-13X (Zeochem Co.) \cite{mofarahi2015experimental}, which have a microporous volume of 0.21 cm$^3$/g and a BET surface of 164.3 m$^2$/g (\cite{zhang2010adsorption}). The measurements were performed volumetrically at 303K and 323K over a pressure ranging from 105 KPa to 705 KPa. Overall, the new model overestimates the mixture adsorption by 5.95\%, while the pure isotherms are underestimated by 2.62\%.

Table \ref{Table12} gives the mean error between the prediction of both approaches.

\begin{table}[!htb]
\centering
\small
\begin{tabular}{llccccc}
							&	& \multicolumn{5}{c}{Mean error (\%)} \\\cline{3-7}
							&	& \multicolumn{2}{c}{Std MPTA} && \multicolumn{2}{c}{New MPTA} \\\cline{3-4}\cline{6-7}
System				&		& N$_{ex}^i$	& S$_{CH_4/N_2}$	&& N$_{ex}^i$	& S$_{CH_4/N_2}$	 \\\hline
CH$_4$/N$_2$ 303K			& CH$_4$ component				& 7.87 	& ---		&& 7.99	& --- \\
							& N$_2$ component				& 6.06 	& 14.31	&& 3.37	& 9.52 \\
							& Mixture							& 6.20 	& ---		&& 6.20	& --- \\\hline
CH$_4$/N$_2$ 323K			& CH$_4$ component				& 13.11 	& ---		&& 8.68 	& --- \\
							& N$_2$ component				& 5.30 	& 17.39	&& 5.64	& 10.32 \\
							& Mixture							& 9.45 	& ---		&& 8.89	& ---	 \\\hline
\multicolumn{2}{l}{Overall error:}									& 7.88	& 15.71	&& 6.71	& 9.88 \\\hline
\multicolumn{2}{l}{Overall increased performance:} 					& --- 	& ---		&& 14.85	& 37.11  \\\hline
\multicolumn{7}{l}{33 experimental data points.}
\end{tabular}
\caption{\label{Table12}\textsl{Comparison of standard and new MPTA models  on Zeolite-13X at 303K and 323K, and pressure up  to 700 KPa}}
\end{table}

\FloatBarrier
\subsection{Zeolite-13X (Hefti)}

Binary mixtures of CO$_2$ and N$_2$ are considered on Zeolite-13X (Zeochem Co.) \cite{hefti2015adsorption}, which have a microporous volume of 0.21 cm$^3$/g and a BET surface of 164.3 m$^2$/g (\cite{zhang2010adsorption}). The measurements were performed volumetrically at 298K and 318K over a pressure ranging from 115 KPa to 1020 KPa. Overall, the new model underestimates the mixture adsorption by 16.65\%, while the pure isotherms are underestimated by 0.57\%.

Table \ref{Table13} gives the mean error between the prediction of both approaches.

\begin{table}[!htb]
\centering
\small
\begin{tabular}{llccccc}
							&	& \multicolumn{5}{c}{Mean error (\%)} \\\cline{3-7}
							&	& \multicolumn{2}{c}{Std MPTA} && \multicolumn{2}{c}{New MPTA} \\\cline{3-4}\cline{6-7}
System				&		& N$_{ex}^i$	& S$_{CO_2/N_2}$	&& N$_{ex}^i$	& S$_{CO_2/N_2}$	 \\\hline
CO$_2$/N$_2$ 298K			& CO$_2$ component				& 5.93 	& ---			&& 4.56	& --- \\
							& N$_2$ component				& 58.20 	& ---$^\dag$	&& 53.12	& ---$^\dag$ \\
							& Mixture							& 3.35 	& ---			&& 2.24	& --- \\\hline
CO$_2$/N$_2$ 318K			& CO$_2$ component				& 6.46 	& ---			&& 6.07 	& --- \\
							& N$_2$ component				& 59.97 	& ---$^\dag$	&& 49.07	& ---$^\dag$ \\
							& Mixture							& 4.15 	& ---			&& 4.16	& ---	 \\\hline
\multicolumn{2}{l}{Overall error:}									& 22.96	& ---			&& 19.88	& --- \\\hline
\multicolumn{2}{l}{Overall increased performance:} 					& --- 	& ---			&& 13.41	& ---  \\\hline
\multicolumn{7}{l}{11 experimental data points.} \\
\multicolumn{7}{l}{\footnotesize $^\dag$ Error on selectivity over 100\% due to large error on the least adsorbed component.}
\end{tabular}
\caption{\label{Table13}\textsl{Comparison of standard and new MPTA models on Zeolite-13X at 298K and 318K, and pressure up  to 1 MPa}}
\end{table}

\FloatBarrier
\subsection{Zeolite-ZSM-5 (Hefti)}

Binary mixtures of CO$_2$ and N$_2$ are considered on Zeolite-ZSM-5 (Zeochem Co.) \cite{hefti2015adsorption}, which have a microporous volume of 0.155 cm$^3$/g and a BET surface from 264 to 312.4 m$^2$/g (\cite{sang2004difference}). The measurements were performed volumetrically at 298K and 318K over a pressure ranging from 120 KPa to 1010 KPa. Overall, the new model underestimates the mixture adsorption by 7.51\%, while the pure isotherms are underestimated by 1.48\%.

Table \ref{Table14} gives the mean error between the prediction of both approaches.

\begin{table}[!htb]
\centering
\small
\begin{tabular}{llccccc}
							&	& \multicolumn{5}{c}{Mean error (\%)} \\\cline{3-7}
							&	& \multicolumn{2}{c}{Std MPTA} && \multicolumn{2}{c}{New MPTA} \\\cline{3-4}\cline{6-7}
System				&		& N$_{ex}^i$	& S$_{CO_2/N_2}$	&& N$_{ex}^i$	& S$_{CO_2/N_2}$	 \\\hline
CO$_2$/N$_2$ 298K			& CO$_2$ component				& 2.08 	& ---			&& 2.36	& --- \\
							& N$_2$ component				& 44.80 	& 106.17		&& 26.40	& 43.91 \\
							& Mixture							& 2.09 	& ---			&& 1.21	& --- \\\hline
CO$_2$/N$_2$ 318K			& CO$_2$ component				& 2.27 	& ---			&& 2.00 	& --- \\
							& N$_2$ component				& 28.35 	& 50.85		&& 19.79	& 26.24 \\
							& Mixture							& 3.24 	& ---			&& 2.23	& ---	 \\\hline
\multicolumn{2}{l}{Overall error:}									& 13.48	& 74.90		&& 8.87	& 33.92 \\\hline
\multicolumn{2}{l}{Overall increased performance:} 					& --- 	& ---			&& 34.20	& 54.71  \\\hline
\multicolumn{7}{l}{23 experimental data points.} 
\end{tabular}
\caption{\label{Table14}\textsl{Comparison of standard and new MPTA models  on Zeolite-ZSM-5 at 298K and 318K, and pressure up  to 1 MPa}}
\end{table}

\FloatBarrier
\subsection{Zeolite-NaX (Belmabkhout)}

Binary mixtures of CO$_2$ and CO are considered on Zeolite-NaX, which have a microporous volume of 0.283 cm$^3$/g and a BET surface of 685 m$^2$/g (\cite{belmabkhout2007}). The measurements were performed volumetrically at 323K and 373K at a pressure of 100 KPa. Overall, the new model underestimates the mixture adsorption by 6.15\%, while the pure isotherms are overestimated by 0.65\%.

Table \ref{Table15} gives the mean error between the prediction of both approaches.

\begin{table}[!htb]
\centering
\small
\begin{tabular}{llccccc}
							&	& \multicolumn{5}{c}{Mean error (\%)} \\\cline{3-7}
							&	& \multicolumn{2}{c}{Std MPTA} && \multicolumn{2}{c}{New MPTA} \\\cline{3-4}\cline{6-7}
System				&		& N$_{ex}^i$	& S$_{CO_2/CO}$	&& N$_{ex}^i$	& S$_{CO_2/CO}$	 \\\hline
CO$_2$/CO 323K				& CO$_2$ component				& 22.33 	& 43.22		&& 24.51	& 36.82 \\
							& CO component					& 26.22 	& ---			&& 19.47	& --- \\
							& Mixture							& 24.62 	& ---			&& 21.52	& --- \\\hline
CO$_2$/CO 373K				& CO$_2$ component				& 19.13 	& 33.52		&& 19.66 & 33.63 \\
							& CO component					& 22.04 	& ---			&& 22.06	& --- \\
							& Mixture							& 16.39 	& ---			&& 16.80	& ---	 \\\hline
\multicolumn{2}{l}{Overall error:}									& 21.59	& 36.75		&& 20.28	& 34.69 \\\hline
\multicolumn{2}{l}{Overall increased performance:} 					& --- 	& ---			&& 6.07	& 5.61  \\\hline
\multicolumn{7}{l}{3 experimental data points.} 
\end{tabular}
\caption{\label{Table15}\textsl{Comparison of standard and new MPTA models  on Zeolite-NaX at 323K and 373K under 1 Bar}}
\end{table}

\FloatBarrier
\subsection{Mordenite (Talu)}

Binary and ternary mixtures adsorption of CO$_2$, H$_2$S and C$_3$H$_8$ are studied on hydrogen mordenite (Norton Company) \cite{talu1986}, which have a BET surface of 400 m$^2$/g (\cite{sagert1972benzene}). The measurements were performed volumetrically at 303K over a pressure ranging from 1 KPa to 61 KPa. Overall, the new model underestimates the mixture adsorption by 23.88\%, while the pure isotherms are overestimated by 0.59\%.

Table \ref{Table16} gives the mean error between the prediction of both approaches.

\begin{table}[!htb]
\centering
\small
\begin{tabular}{llccccc}
							&	& \multicolumn{5}{c}{Mean error (\%)} \\\cline{3-7}
							&	& \multicolumn{2}{c}{Std MPTA} && \multicolumn{2}{c}{New MPTA} \\\cline{3-4}\cline{6-7}
System				&			& N$_{ex}^i$	& Select			&& N$_{ex}^i$	& Select	 \\\hline
CO$_2$/H$_2$S				& CO$_2$ component		& 36.12 	& ---			&& 51.08		& ---	  \\
							& H$_2$S component		& 7.33	& 72.78 		&& 10.37 	& 145.12 \\
							& Mixture					& 6.32 	& ---			&& 2.20		& ---  \\\hline
C$_3$H$_8$/CO$_2$			& C$_3$H$_8$ component	& 22.62 	& 100.45		&& 16.70		& 84.50	 \\
							& CO$_2$ component		& 49.83 	& ---			&& 47.31 	& ---	\\
							& Mixture					& 16.85 	& ---			&& 21.76 	& --- 	\\\hline
C$_3$H$_8$/H$_2$S			& C$_3$H$_8$ component	& 47.15 	& ---			&& 45.10		& ---	 \\
							& H$_2$S component		& 17.42 	& ---$^\dag$	&& 15.93 	& ---$^\dag$	\\
							& Mixture					& 20.42 	& ---			&& 20.12 	& --- 	\\\hline
CO$_2$/H$_2$S/C$_3$H$_8$	& CO$_2$ component		& 50.59 	& ---			&& 63.64		& ---	  \\
							& H$_2$S component		& 92.01 	& ---$^\dag$	&& 100.81 	& ---$^\dag$ \\
							& C$_3$H$_8$ component	& 82.49 	& ---$^\dag$ 	&& 83.15		& ---$^\dag$ 	 \\
							& Mixture					& 29.42 	& ---			&& 31.45		& --- \\\hline
\multicolumn{2}{l}{Overall error:}							& 37.94	& 86.62		&& 40.55		& 114.81 \\
\multicolumn{2}{l}{Overall increased performance:} 			& --- & ---				&& -6.88 		& -32.54  \\\hline
\multicolumn{7}{l}{36 experimental data points.} \\
\multicolumn{7}{l}{\footnotesize $^\dag$ Error on selectivity over 100\% due to large error on the least adsorbed component.}
\end{tabular}
\caption{\label{Table16}\textsl{Comparison of standard and new MPTA models on H-Mordenite at 303K and pressure from 1 to 61 KPa}}
\end{table}

\FloatBarrier
\section{Conclusion}

A new approach to the Multicomponent Potential Theory of Adsorption was presented in which individual fitting parameters replaced the commons ones. Specifically, the new approach uses distinct values of the parameters $z_0$ (the limiting microporous volume) and $\beta$ (the heterogeneity parameter) for the model pure gases fits. In the standard MPTA model, those parameters are shared by all the pure gases, which generated the coupling of pure gases. In this new interpretation of the model, there are individual parameters for each pure gases considered. This interpretation implies more fitting parameters (3$M$ parameters instead of $M$+2) but is nevertheless easier to understand and adjust because the model decomposed into $M$ individual three parameters fit. The objective pursued is the ability to predict mixture adsorption without any experimental measurements by extrapolating parameters from one adsorbent to another. Under that scope, the independence of the components is a crucial step.\\

To ensure consistency of the experimental data to be compared with, the relative experimental uncertainty was calculated, and data with larger than 25\% relative uncertainties were omitted from the fits. For datasets without given experimental uncertainty, a relative uncertainty of 1\% on total excess adsorption and the smallest component molar fraction was assumed to evaluate relative uncertainties. Those uncertainties are representative of what is found in the literature. Finally, after testing over 500 experimental mixture data, the presented interpretation performed 4.67\% better than the usual model, which gives a mean error of 6.97\% for total mixture excess adsorption, and an overall combined mean error of 15.30\% for component and mixture excess adsorption.  

\section*{Acknowledgements}

The authors would like to thanks the Natural Sciences and Engineering Research Council of Canada and the Savannah River National Laboratory for financial support.

\FloatBarrier
\bibliographystyle{ieeetr}
\bibliography{biblio.bib}

\end{document}